\documentclass[10pt,a4paper]{article}
\usepackage[dvips]{color}
\usepackage{epsfig}
\usepackage{amsmath}
\usepackage{graphicx}
\usepackage{authblk}
\usepackage{amssymb,amsmath}
\usepackage{amsmath}

\begin{document}
\title{\bf Interacting two-component fluid models with varying EoS parameter}
\author{{M. Khurshudyan$^{a}$ \thanks{Email: khurshudyan@yandex.ru},\hspace{2mm} B. Pourhassan$^{b, c}$ \thanks{Email: b.pourhassan@umz.ac.ir},\hspace{2mm} and E.O. Kahya$^{c}$\thanks{Email: eokahya@itu.edu.tr}}\\
$^{a}${\small {\em Department of Theoretical Physics, Yerevan State
University, 1 Alex Manookian, 0025, Yerevan, Armenia}}\\
$^{b}${\small {\em Department of Physics, Damghan
University, Damghan, Iran}}\\
$^{c}${\small {\em Physics Department, Istanbul Technical University, Istanbul, Turkey}}} \maketitle
\begin{abstract}
In this paper, we consider Universe filled with two-component fluid. We study two different models. In the first model we assume barotropic fluid with the linear
equation of state as the first component of total fluid. In the second model we assume Van der Waals gas as the first component of total fluid. In both models, the second component assumed generalized ghost dark energy. We consider also interaction between components and discuss, numerically, cosmological quantities for two different parametrizations of EoS which varies with time. We consider this as a toy model of our Universe. We fix parameters of the model by using generalized second law of thermodynamics. Comparing our results with some observational data suggests interacting barotropic fluid with EoS parameter $\omega(t)=\omega_{0}\cos(tH)+\omega_{1}t\frac{\dot{H}}{H}$ and generalized ghost dark energy as an appropriate model to describe our Universe.
\end{abstract}
\section{Introduction}
Among the different interesting problems of theoretical cosmology, an intriguing problem concerning present epoch of Universe is still open. Observational data indicate that our Universe has accelerated expansion. Einstein general relativity proposes concept of dark energy to describe an accelerated expansion of Universe. Dark energy is a fluid with negative pressure and positive energy density therefore has negative equation of state (EoS) parameter. There are various models to describe dark energy. One of them is cosmological constant model so it is a perfect fit to the dark energy data. But there are two problems called fine-tuning and coincidence [1] which suggest alternative models to describe dark energy such as quintessence [2] and k-essence [3] which obtained by modifying the right hand side of Einstein equation (modified matter models). On the other hand one can modifies the left hand side of Einstein equation and obtain modified gravity such as $f(r)$ gravities [4, 5, 6]. Modifications of these types provide an origin of a fluid (energy density and pressure) identified with dark energy. These kinds of fluids obviously have geometrical origin. A possibility to have accelerated expansion contributed from geometry were considered even before proposed modifications and this approach could be accepted as true-like feeling. But such theories with different forms of modifications still should pass experimental tests, because they contain ghosts, finite-time future singularities e.t.c, which is the base of other theoretical problems.\\
Another interesting models of dark energy which have several attentions from theoretical physics are Chaplygin gas and its extension models [7-17]. These models enable us to study dynamics of dark energy. However, there is still another interesting models such as ghost dark energy models [18] which is used in this paper. In this paper we are interested to the Universe with two-component interacting fluid. Considering two-component fluid instead of single-component fluid help us to fix solution better with observational data. We will consider two models, the first one includes a barotropic fluid with the linear
equation of state as the first component of fluid where the dark energy pressure is given as an explicit function of the density [19]. In the second model we assume that the first component of fluid is a Van der Waals gas [20-22]. This model may describes the transition from a
scalar field dominated epoch to a matter field dominated
period without introducing scalar fields which is an importance of this model. In both models the second component is a generalized ghost dark energy \cite{Cai}. The ghost dark energy model may solve the $U(1)_{A}$ problem in low-energy effective theory of QCD. Actually, the contribution of the ghost fields to the vacuum energy may be regarded as a possible candidate for the dark energy. Also we consider two different parametrization of time-dependent EoS parameter. Already EoS parameter considered as $\omega(t)=\omega_{0}+\omega_{1}t\frac{\dot{H}}{H}$ \cite{Khurshudyan} so we attempt to generalize parametrization of EoS parameter.\\
This paper is organized as the following. In the next section we write field equations and metric of our Universe. Then, in section 3 we introduce our two-component fluid models. In section 4 we consider interacting fluids and discuss about cosmological parameters numerically. Finally in section 5 we give conclusion and discuss about results.
\section{Field equations}
We assume FRW metric for a flat Universe,
\begin{equation}\label{s1}
ds^2=-dt^2+a(t)^2\left(dr^{2}+r^{2}d\Omega^{2}\right).
\end{equation}
Field equations that govern our model are,
\begin{equation}\label{s2}
R^{ij}-\frac{1}{2}Rg^{ij}=-8 \pi G T^{ij},
\end{equation}
which can be reduced to the following Friedmann equations,
\begin{equation}\label{eq: Fridmman vlambda}
H^{2}=\frac{\dot{a}^{2}}{a^{2}}=\frac{8\pi G\rho}{3},
\end{equation}
and,
\begin{equation}\label{eq:fridman2}
\frac{\ddot{a}}{a}=-\frac{4\pi
G(t)}{3}(\rho+3P),
\end{equation}
where $d\Omega^{2}=d\theta^{2}+\sin^{2}\theta d\phi^{2}$, and $a(t)$
represents the scale factor.\\
Energy conservation $T^{;j}_{ij}=0$ reads as,
\begin{equation}\label{eq:conservation}
\dot{\rho}+3H(\rho+P)=0.
\end{equation}
\section{Two-component fluid models}
We assume that our Universe filled with two-component fluid so $\rho$ and $P$ in the equations of the previous section decompose as the following,
\begin{eqnarray}\label{s6}
\rho=\rho_{1}+\rho_{2},\nonumber\\
P=P_{1}+P_{2}.
\end{eqnarray}
Therefore total EoS parameter given by,
\begin{equation}\label{eq:EoSpar}
\omega_{tot}=\frac{P_{1}+P_{2}}{\rho_{1}+\rho_{2}}.
\end{equation}
We assume generalized ghost dark energy as the second component of fluid. Energy density of ghost dark energy reads as \cite{Ghost1, Chao-Jun},
\begin{equation}\label{eq:GDE}
\rho_{GDE}=\theta H,
\end{equation}
where $H$ is Hubble parameter $H=\dot{a}/a$ and $\theta$ is constant parameter of the model, which should be determined. A generalization of the model also was proposed for which energy density reads as \cite{Cai},
\begin{equation}\label{eq:GDEgen}
\rho_{GGDE}=\theta H+\xi H^{2},
\end{equation}
where $\theta$ and $\xi$ are constant parameters of the model. We mentioned this model as generalized ghost dark energy (GGDE). Recently, idea of varying ghost dark energy \cite{Khurshudyans} were proposed, which from our opinion still should pass long way with different modifications.
Therefore we will set $\rho_{2}=\rho_{GGDE}$.\\ About the first component of fluid we consider two different models which introduced in the next subsections.
\subsection{The first model}
In the first model we consider barotropic fluid as the first component of fluid which has the following linear
equation of state,
\begin{equation}\label{eq:barfluid}
P_{b}=\omega(t) \rho_{b},
\end{equation}
while it is possible to choose another form of EoS [19] such as quadratic.
One of the important properties of barotropic fluids is that
the speed of sound $C_{s}^{2}=dP/d\rho$, does not have to
equal the speed of light just like to quintessence models.\\
Indeed, the idea of a Universe filled with two-component fluid including barotropic fluid and dark energy already given by the Ref. [33] under assumption of quasiexponential scale factor. Here, we consider two different parametrization as the following.
\subsubsection{Parametrization $I$}
In the first parametrization we choose the following relation,
\begin{equation}\label{eq:omega1}
\omega(t)=q\omega_{0}+\omega_{1}t\frac{\dot{H}}{H},
\end{equation}
where $\omega_{0}$ and $\omega_{1}$ are positive constants, and  $q$ is deceleration parameter given by,
\begin{equation}\label{12}
q=-1-\frac{\dot{H}}{H^{2}}.
\end{equation}
In the plots of the Fig. 1 we draw Hubble expansion, total equation of state and deceleration parameters of the parametrization $I$ of the first model.
\begin{figure}[h!]
 \begin{center}$
 \begin{array}{cccc}
\includegraphics[width=50 mm]{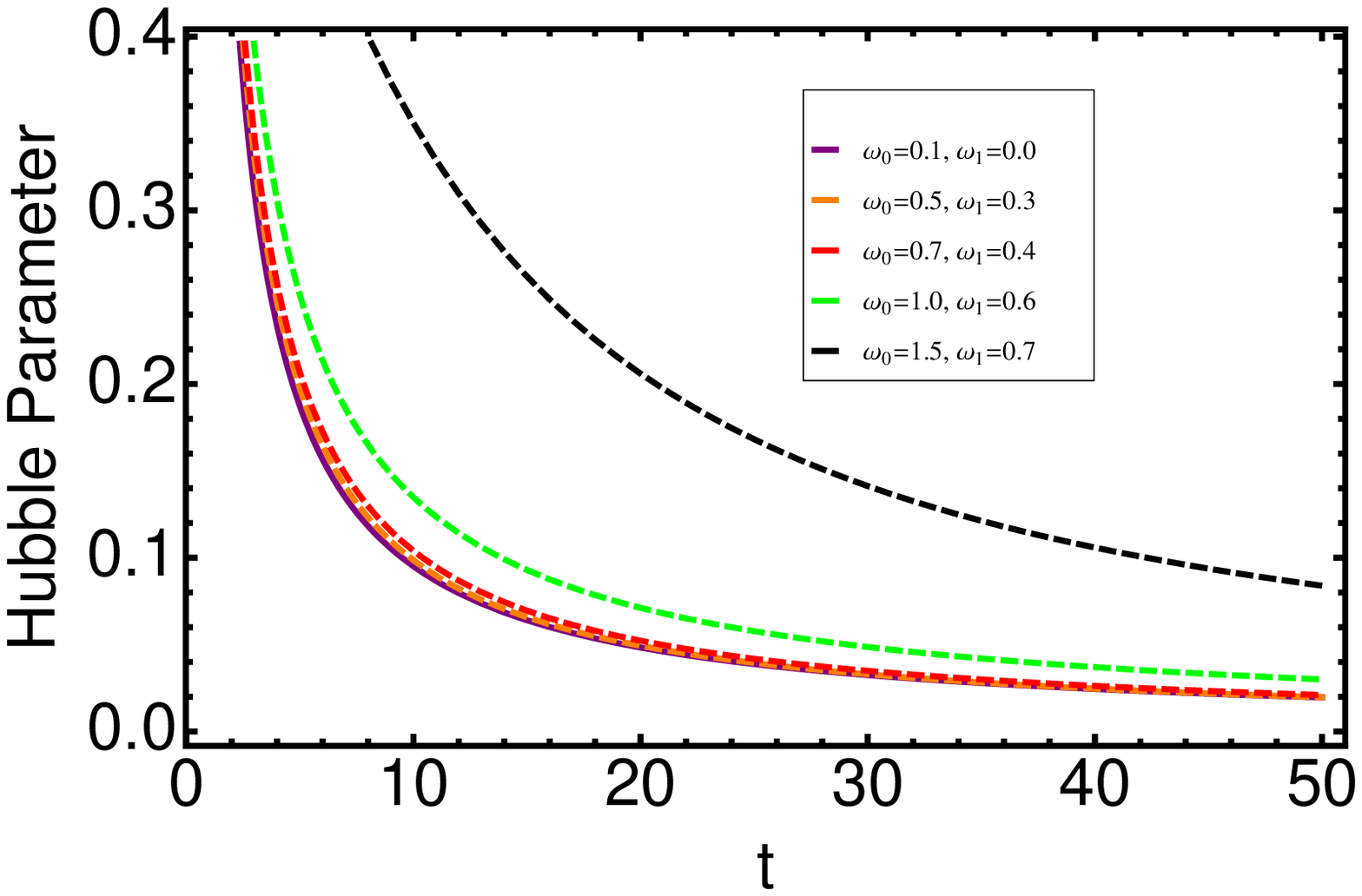} & \includegraphics[width=50 mm]{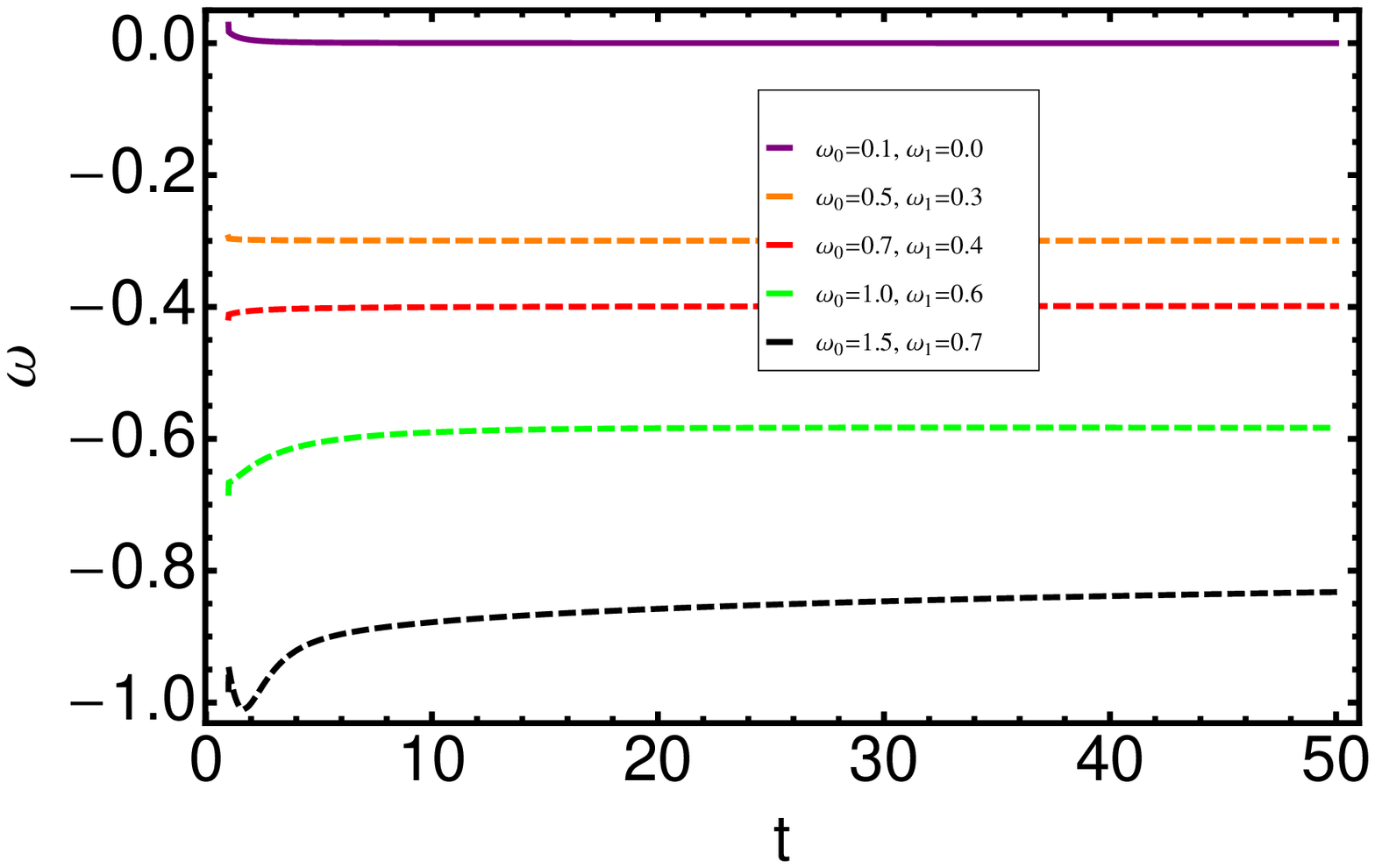} & \includegraphics[width=50 mm]{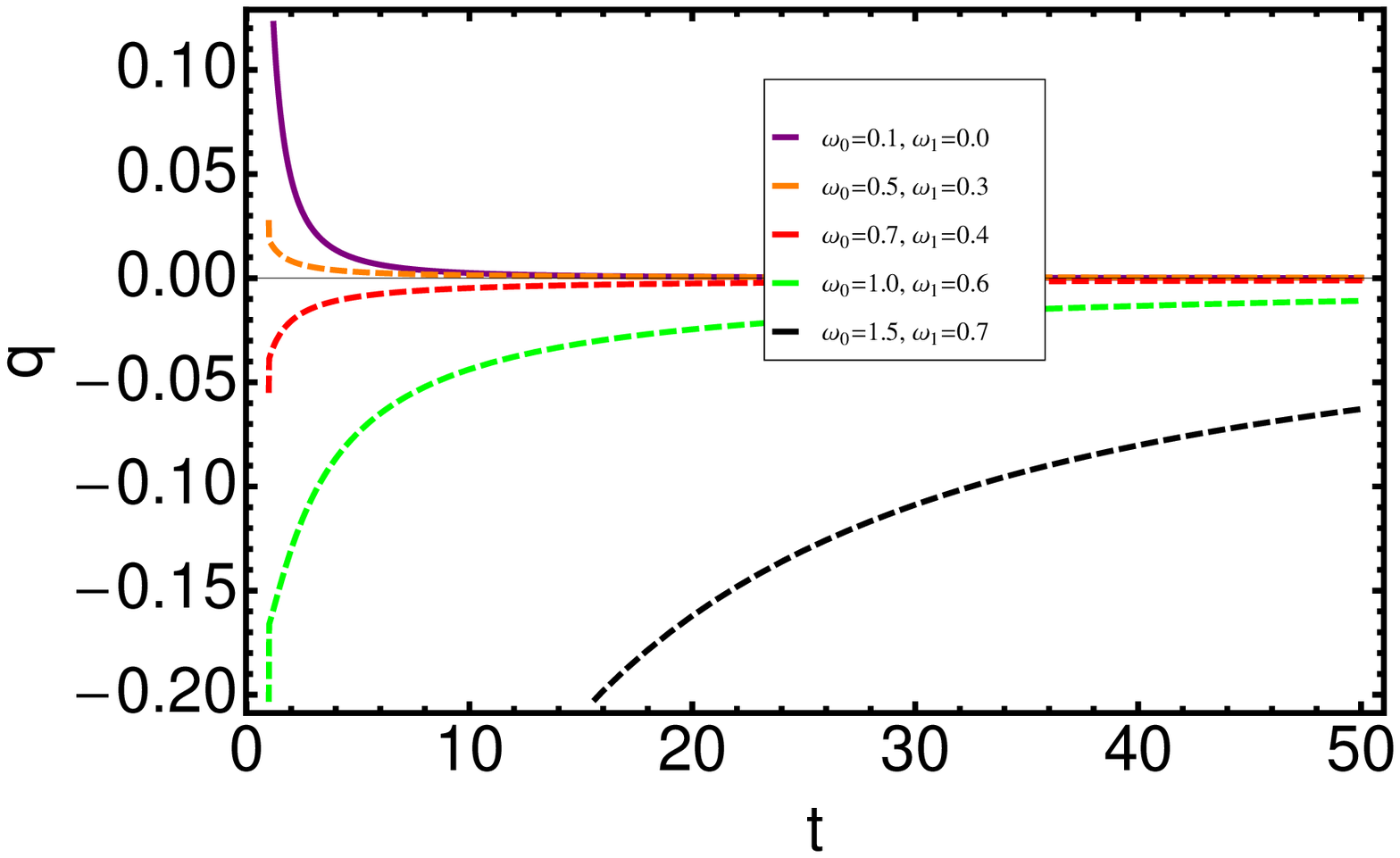}
 \end{array}$
 \end{center}
\caption{Cosmological parameters $H$, $\omega$ and $q$ against $t$ corresponds to a Universe with a barotropic fluid and parametrization $I$.}
 \label{fig:1}
\end{figure}

\subsubsection{Parametrization $II$}
In the second parametrization we choose the following relation,
\begin{equation}\label{eq:omega2}
\omega(t)=\omega_{0}\cos(tH)+\omega_{1}t\frac{\dot{H}}{H},
\end{equation}
In the plots of the Fig. 2 we draw Hubble expansion, total equation of state and deceleration parameters of the parametrization $II$ of the first model.
\begin{figure}[h!]
 \begin{center}$
 \begin{array}{cccc}
\includegraphics[width=50 mm]{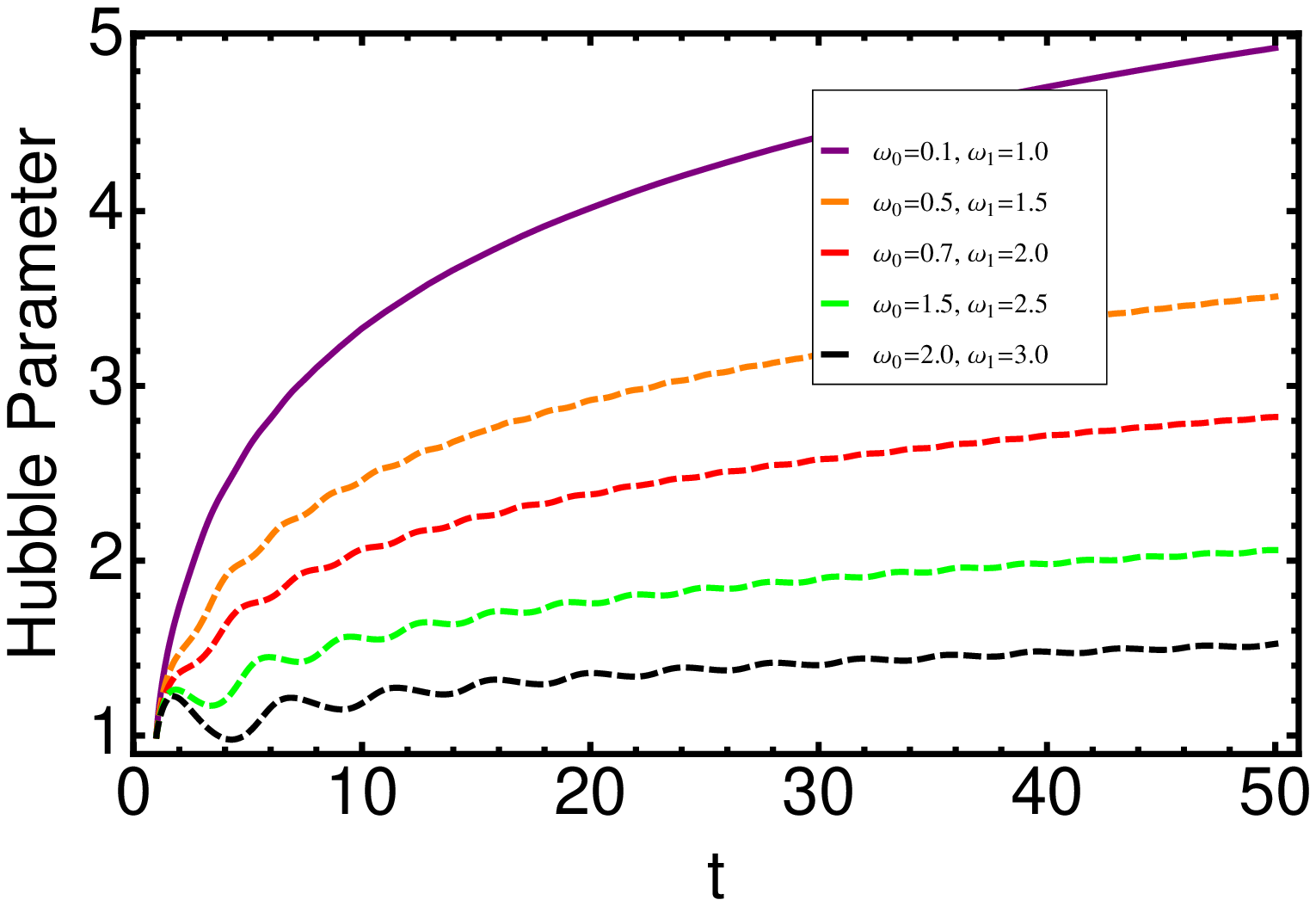} & \includegraphics[width=50 mm]{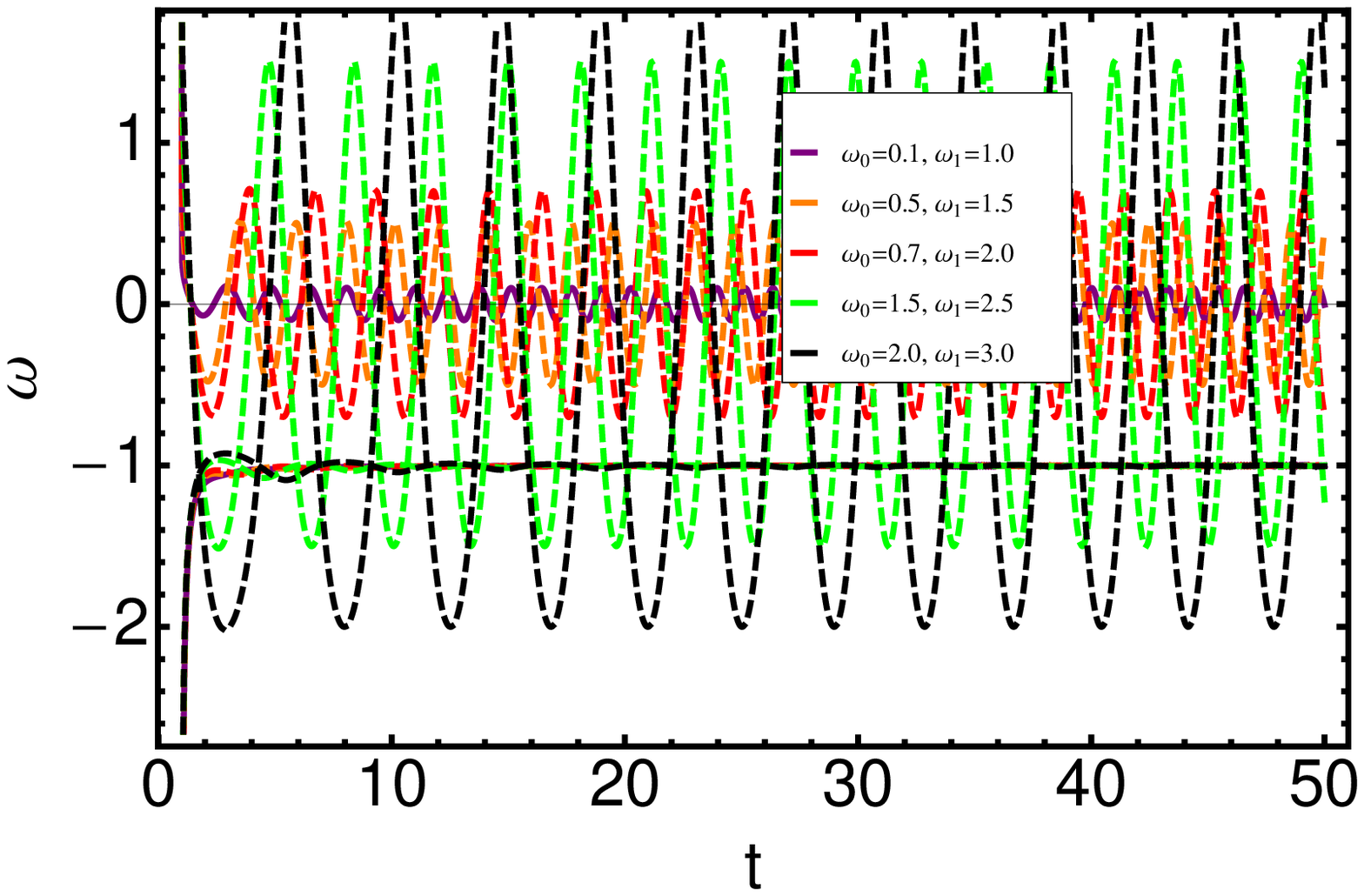} & \includegraphics[width=50 mm]{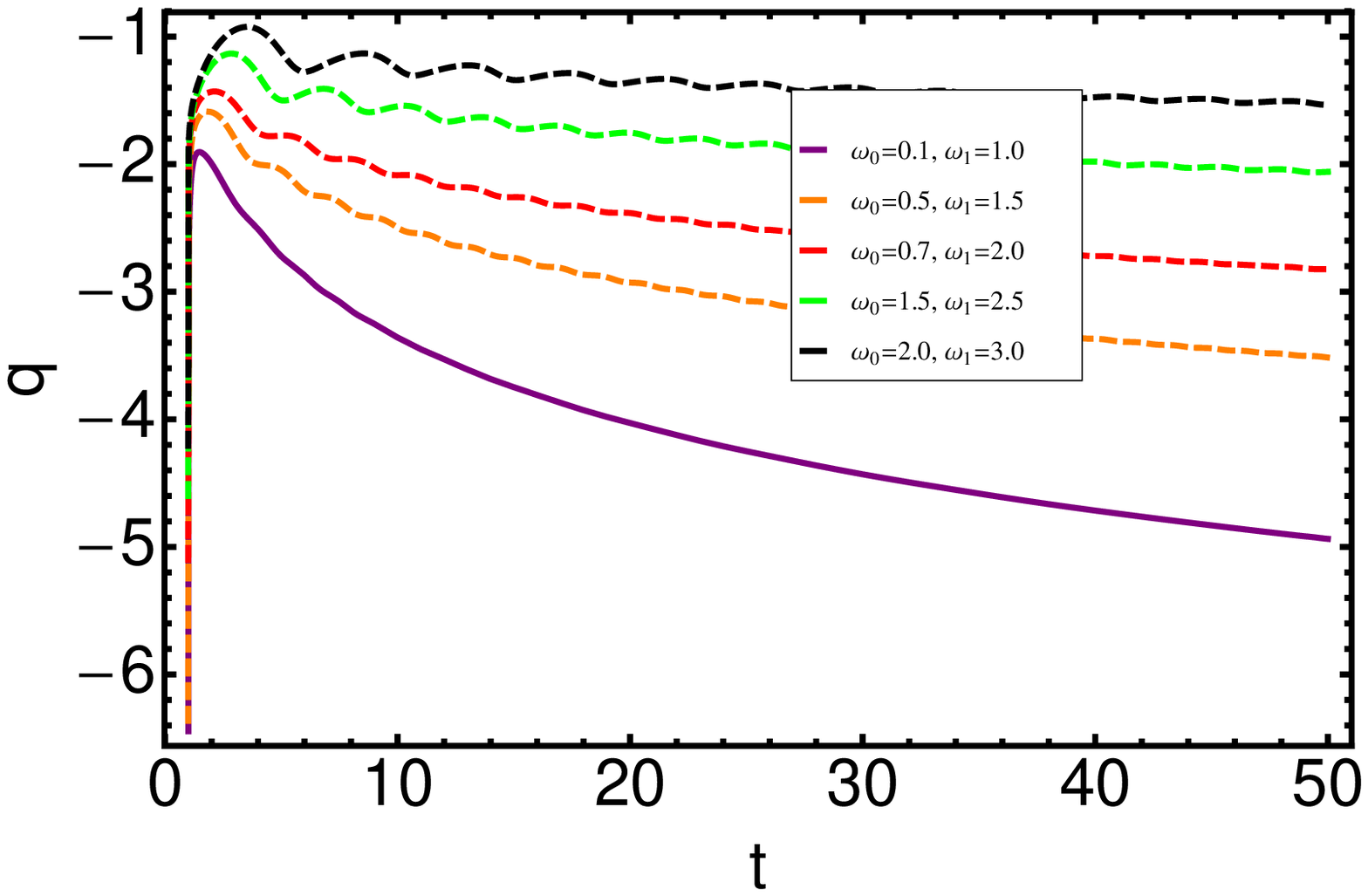}
 \end{array}$
 \end{center}
\caption{Cosmological parameters $H$, $\omega$ and $q$ against $t$ corresponds to a Universe with a barotropic fluid and parametrization $II$.}
 \label{fig:2}
\end{figure}

\subsection{The second model}
In the second model we consider Van der Waals gas as the first component of fluid.
Van der Waals gas (\cite{Khurshudyan1} and references therein) can be understood as a fluid which EoS has the following form,
\begin{equation}\label{eq:waals}
P_{W}=\frac{8\omega(t) \rho_{W} }{ 3-\rho_{W} }-3\rho_{W}^{2},
\end{equation}
where pressure $P_{W}$ and the energy
density $\rho_{W}$ are written in terms of dimensionless reduced
variables. Here we extend EoS introduced in the Ref. [28] to the case of time-dependent $\omega(t)$ instead of a constant parameter connected with a reduced temperature. We use equations (11) and (13) for $\omega(t)$ and obtain the following results corresponding to each parametrization.

\subsubsection{Parametrization $I$}
Using relations (11) and (14) we can draw Hubble expansion, total equation of state and deceleration parameters of the parametrization $I$ in the second model.
\begin{figure}[h!]
 \begin{center}$
 \begin{array}{cccc}
\includegraphics[width=50 mm]{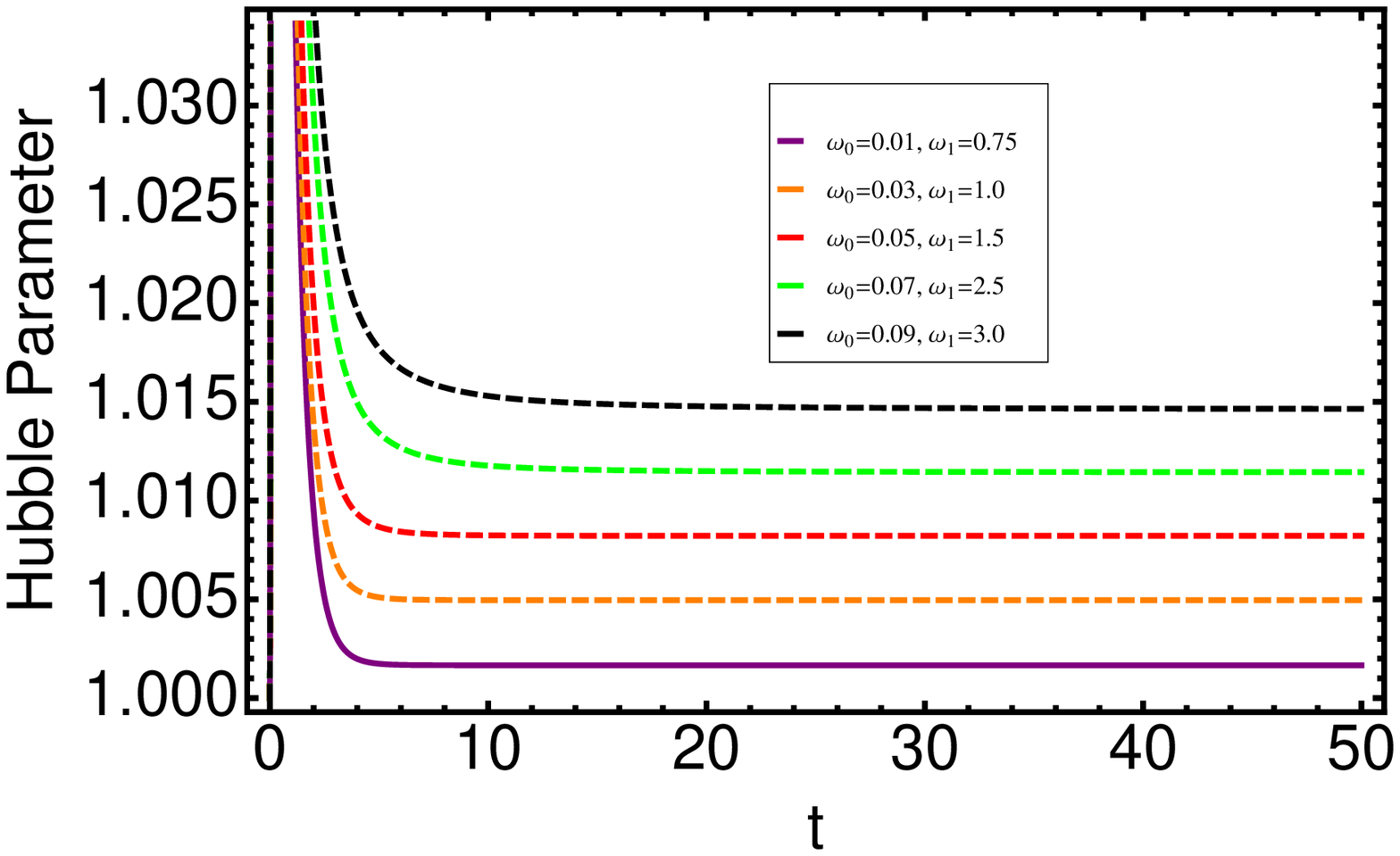} & \includegraphics[width=50 mm]{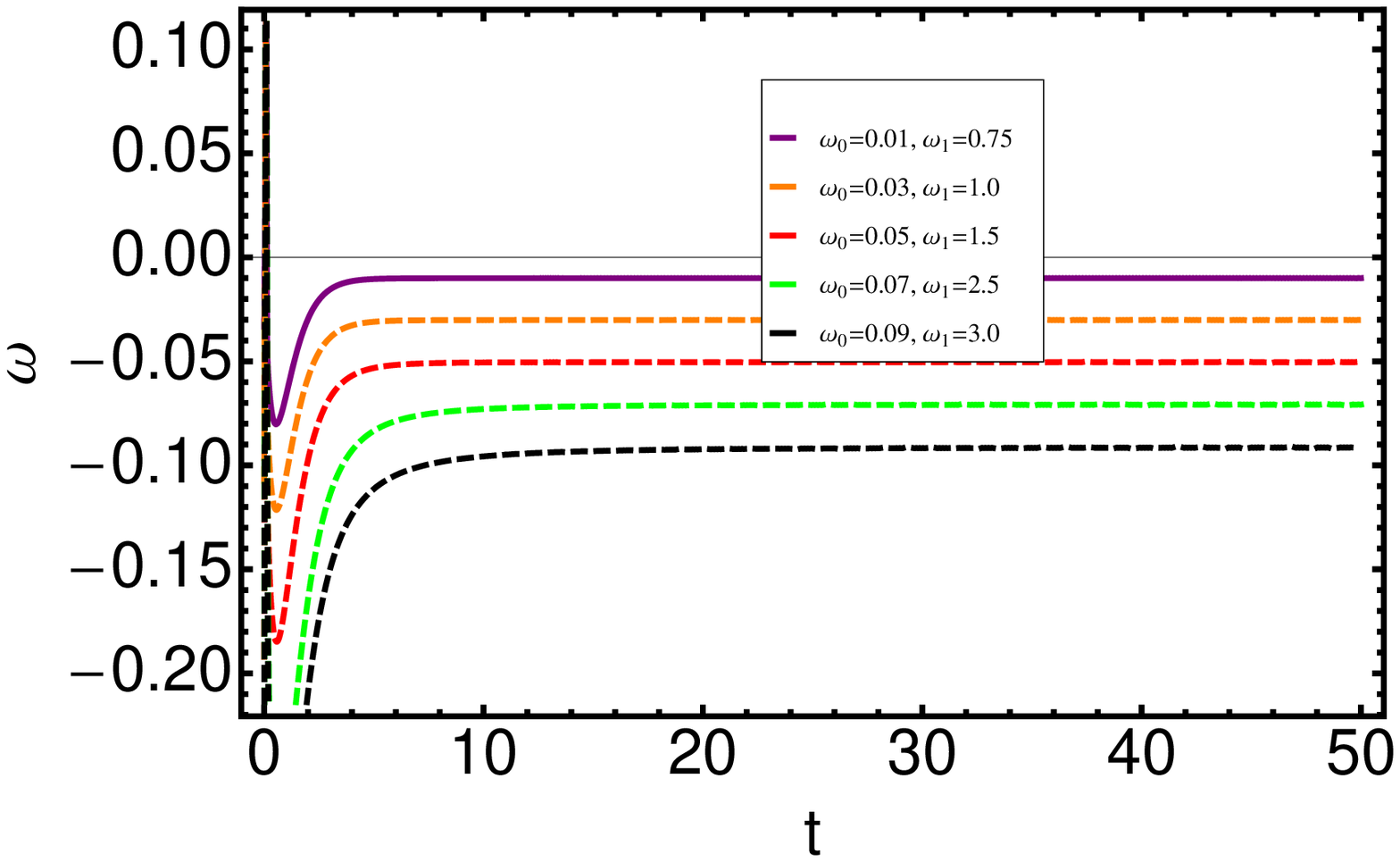} & \includegraphics[width=50 mm]{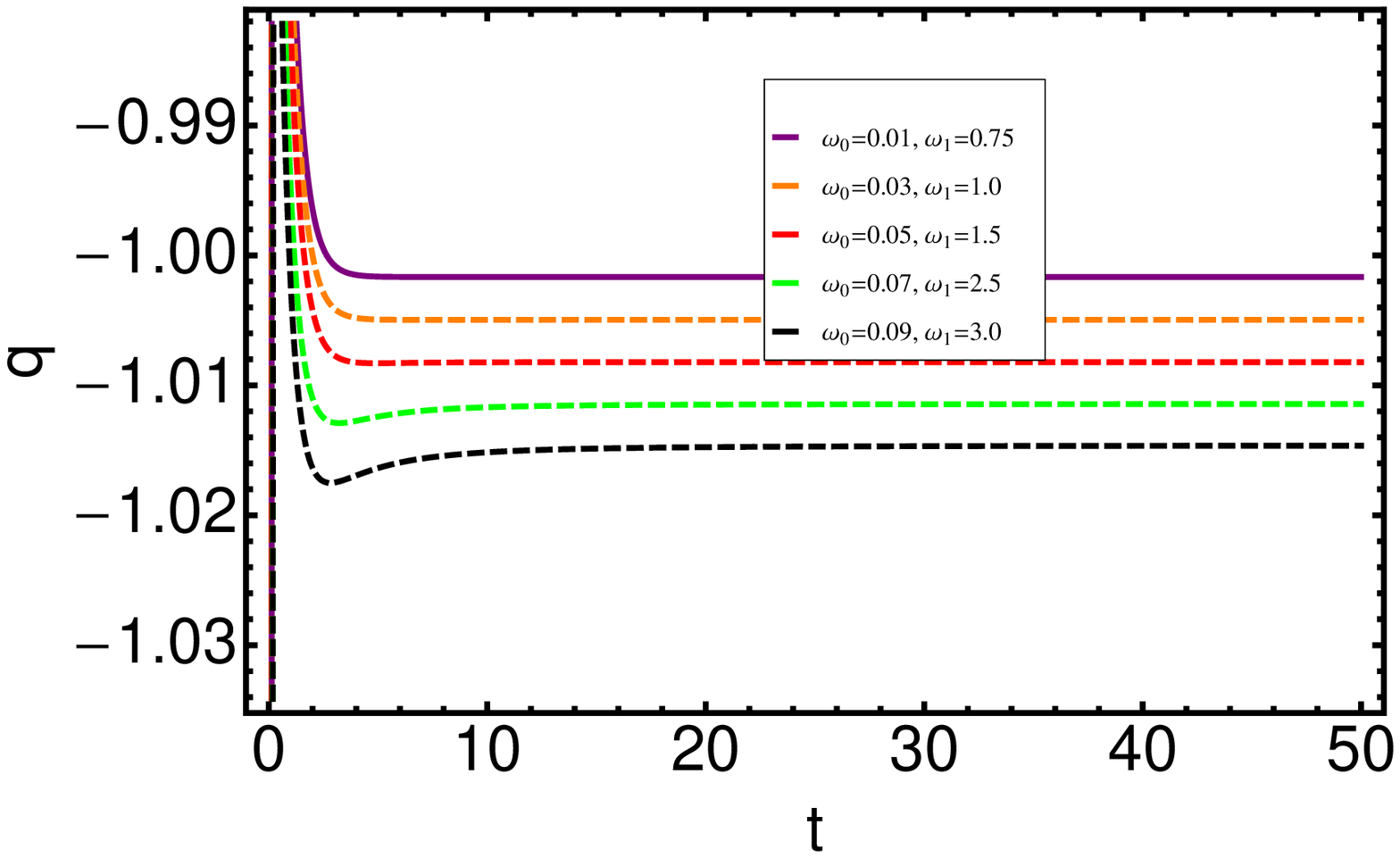}
 \end{array}$
 \end{center}
\caption{Cosmological parameters $H$, $\omega$ and $q$ against $t$ corresponds to a Universe with a Van der Waals gas and parametrization $I$.}
 \label{fig:3}
\end{figure}

\subsubsection{Parametrization $II$}
Using relations (13) and (14) we can draw Hubble expansion, total equation of state and deceleration parameters of the parametrization $II$ in the second model.
\begin{figure}[h!]
 \begin{center}$
 \begin{array}{cccc}
\includegraphics[width=50 mm]{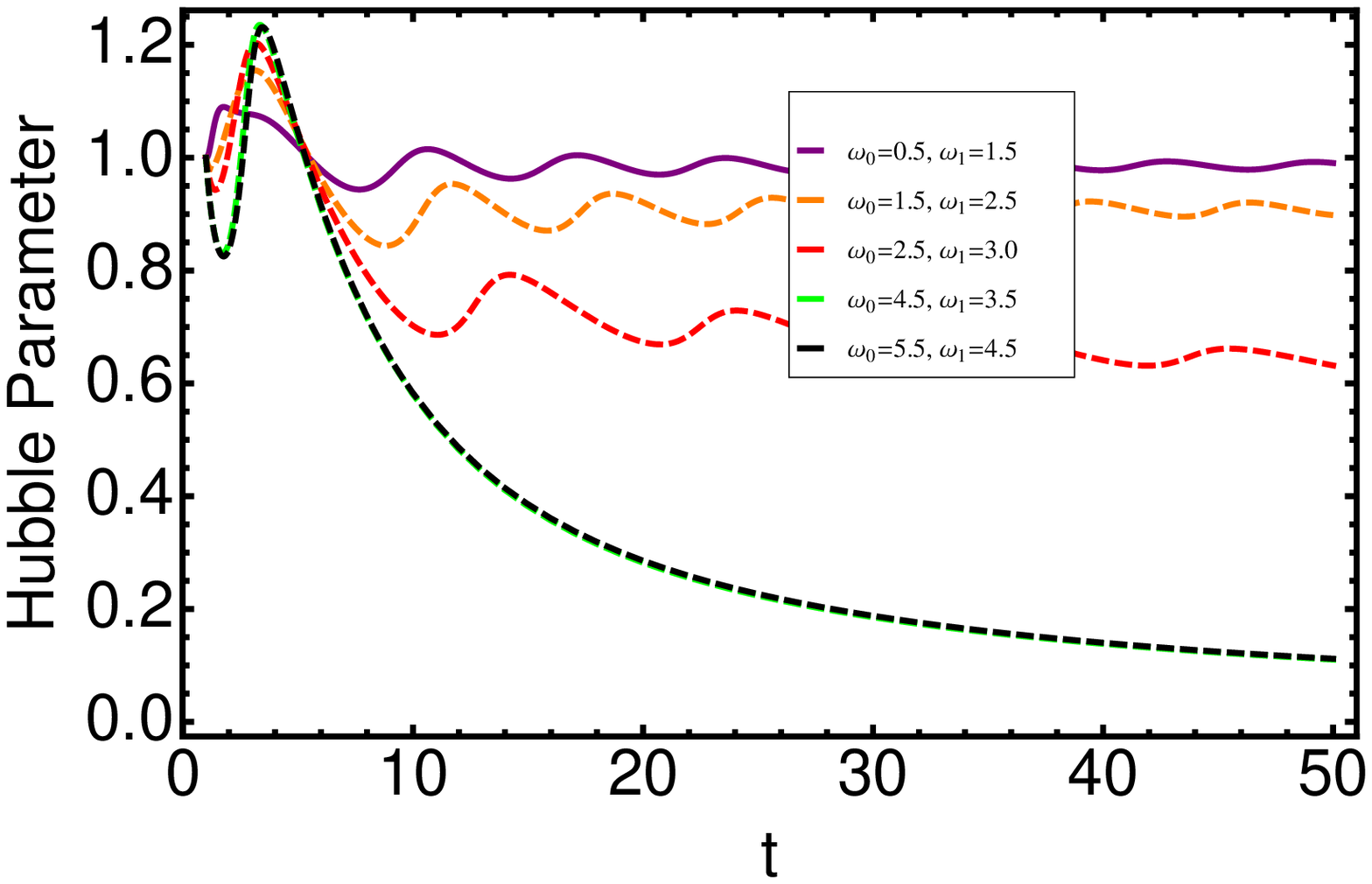} & \includegraphics[width=50 mm]{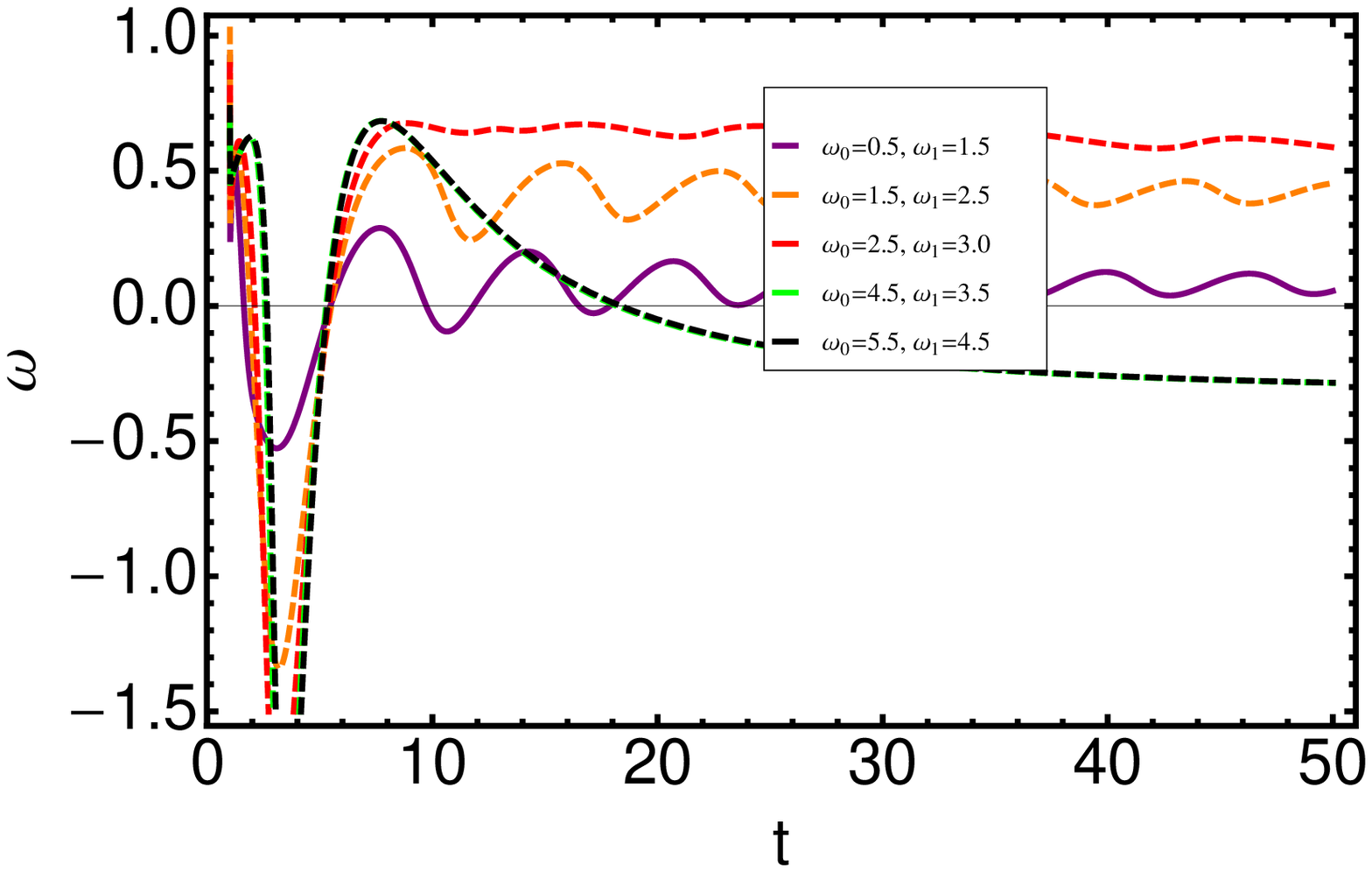} & \includegraphics[width=50 mm]{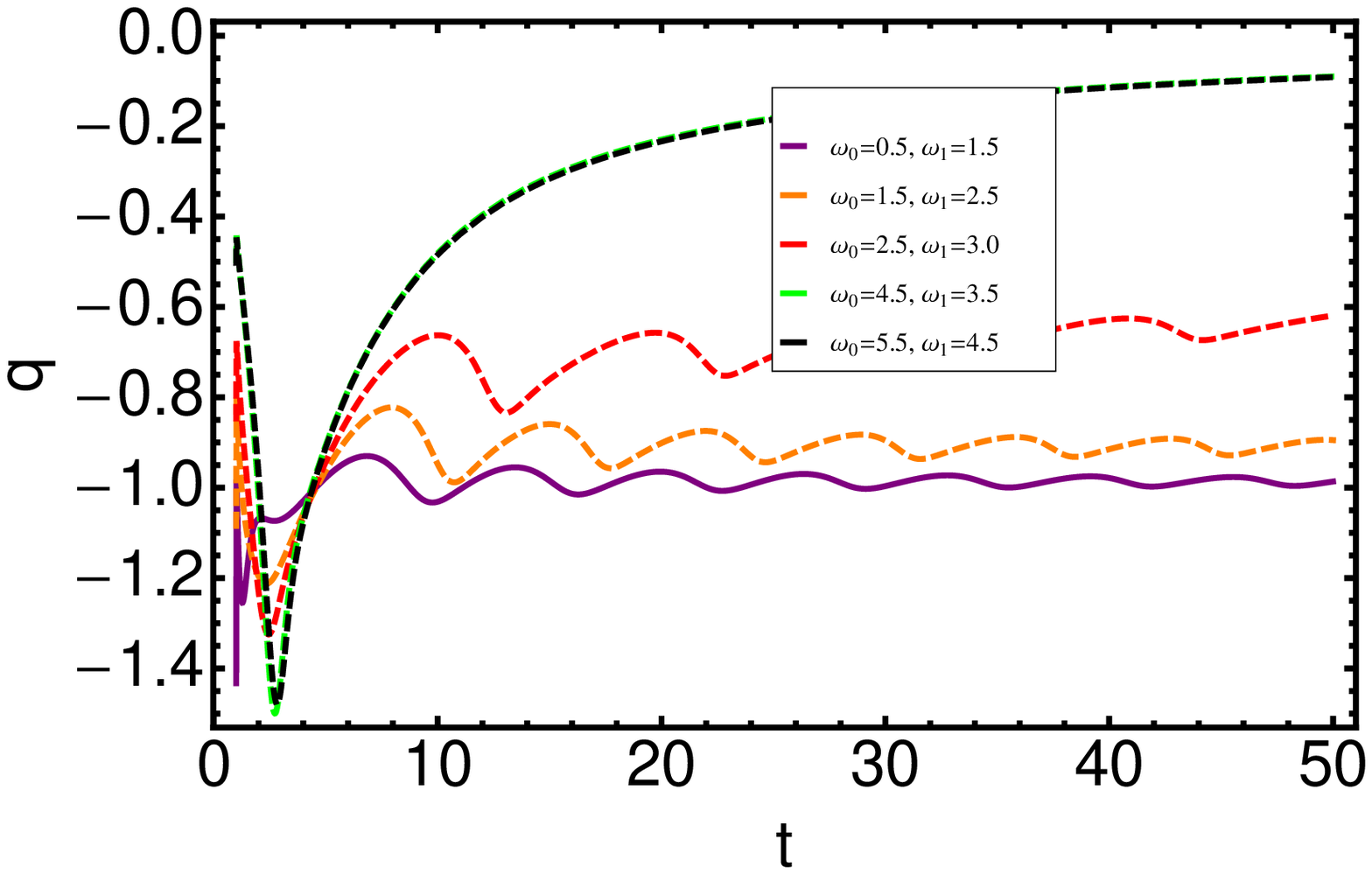}
 \end{array}$
 \end{center}
\caption{Cosmological parameters $H$, $\omega$ and $q$ against $t$ corresponds to a Universe with a Van der Waals gas and parametrization $II$.}
 \label{fig:4}
\end{figure}
\section{Interacting case}
We also consider interaction between components in both models with both parametrization. In that case the equation (5) decomposed as the following,

\begin{eqnarray}\label{eq:conservation}
\dot{\rho}_{1}+3H(\rho_{1}+P_{1})&=&Q,\nonumber\\
\dot{\rho}_{GGDE}+3H(\rho_{GGDE}+P_{GGDE})&=&-Q,
\end{eqnarray}
where $\rho_{1}$ and $P_{1}$ may be density and pressure of barotropic fluid or Van der Waals gas. Also, interaction term assumed as the following,
\begin{equation}\label{eq:int}
Q=3Hb\rho+\gamma \dot{\rho},
\end{equation}
where $\rho=\rho_{1}+\rho_{GGDE}$. In the following subsections we study models and parametrization separately.
\subsection{The first model}
In the first case we consider interacting barotropic fluid and generalized ghost dark energy model and set $\rho_{1}=\rho_{b}$ in the equations (15) and (16).
Dynamics of energy density of barotropic fluid with a varying EoS parameter $\omega(t)$ will be described by the following equation,
\begin{equation}\label{eq:dynamicsb}
(1-\gamma)\dot{\rho}_{b}+3H\rho_{b}(1+\omega(t)-b)-3H^{2}b(\theta+\xi H) -\gamma(\theta \dot{H}+2\xi H\dot{H})=0.
\end{equation}
Pressure of generalized ghost dark energy, after some mathematics, can be presented as follow,
\begin{equation}\label{eq:pressureGDE}
P_{GGDE}=\dot{H} \left ( -\frac{\gamma \theta}{3H}-\frac{2}{3} \gamma \xi \right )-(1+b)(\theta H + \xi H^{2}) -b\rho_{b}-\frac{\gamma \dot{\rho_{b}}}{3H}.
\end{equation}
Results are different for each parametrization which illustrated in the following.
\subsubsection{Parametrization $I$}
If we use relation (11) in the equation (17) then we can study behavior of cosmological parameters $H$, $\omega_{tot}$, $q$ and $\rho_{tot}$ which illustrated in the plots of The Fig. 5.
\begin{figure}[h!]
 \begin{center}$
 \begin{array}{cccc}
\includegraphics[width=50 mm]{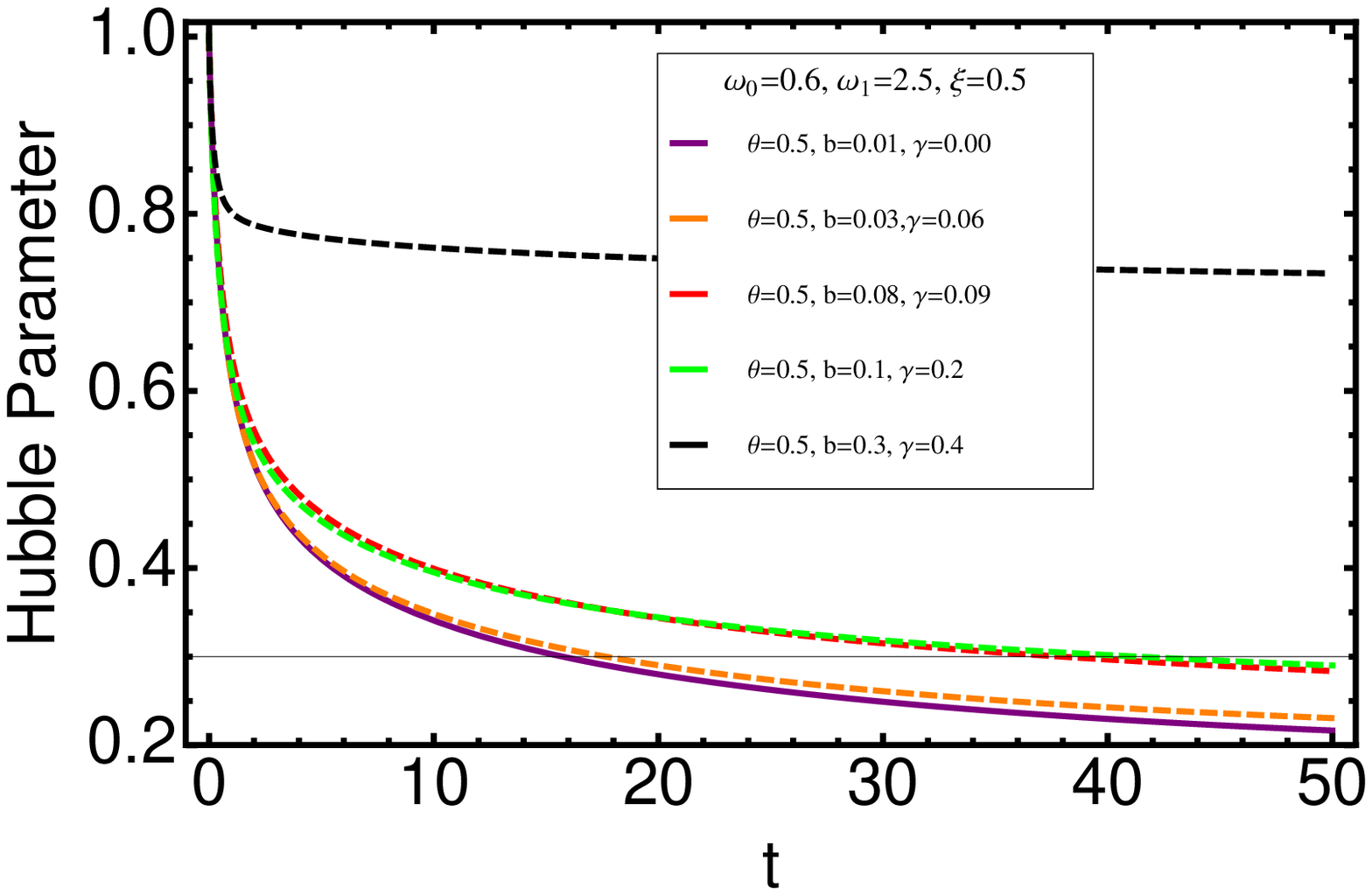} &
\includegraphics[width=50 mm]{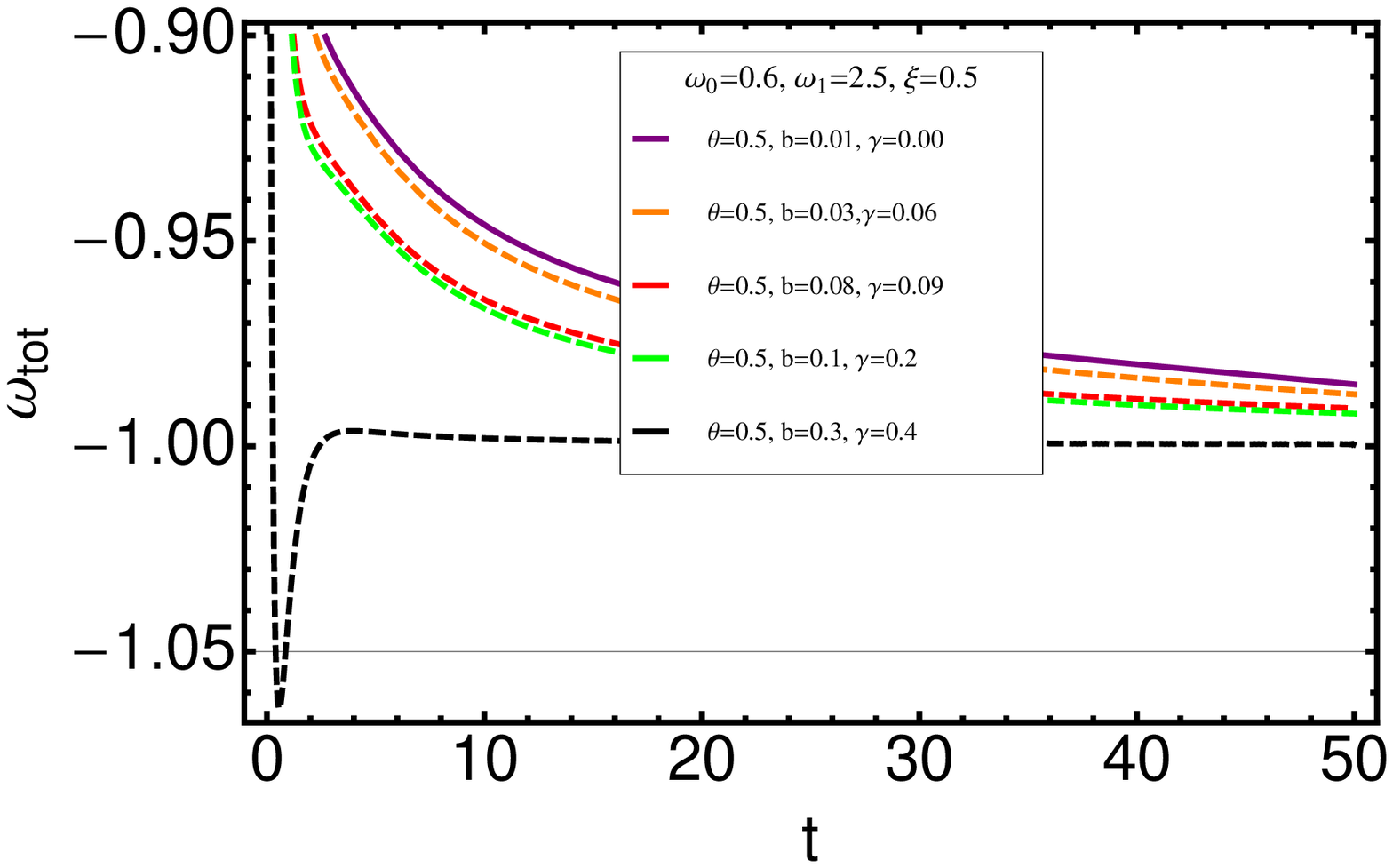}\\
\includegraphics[width=50 mm]{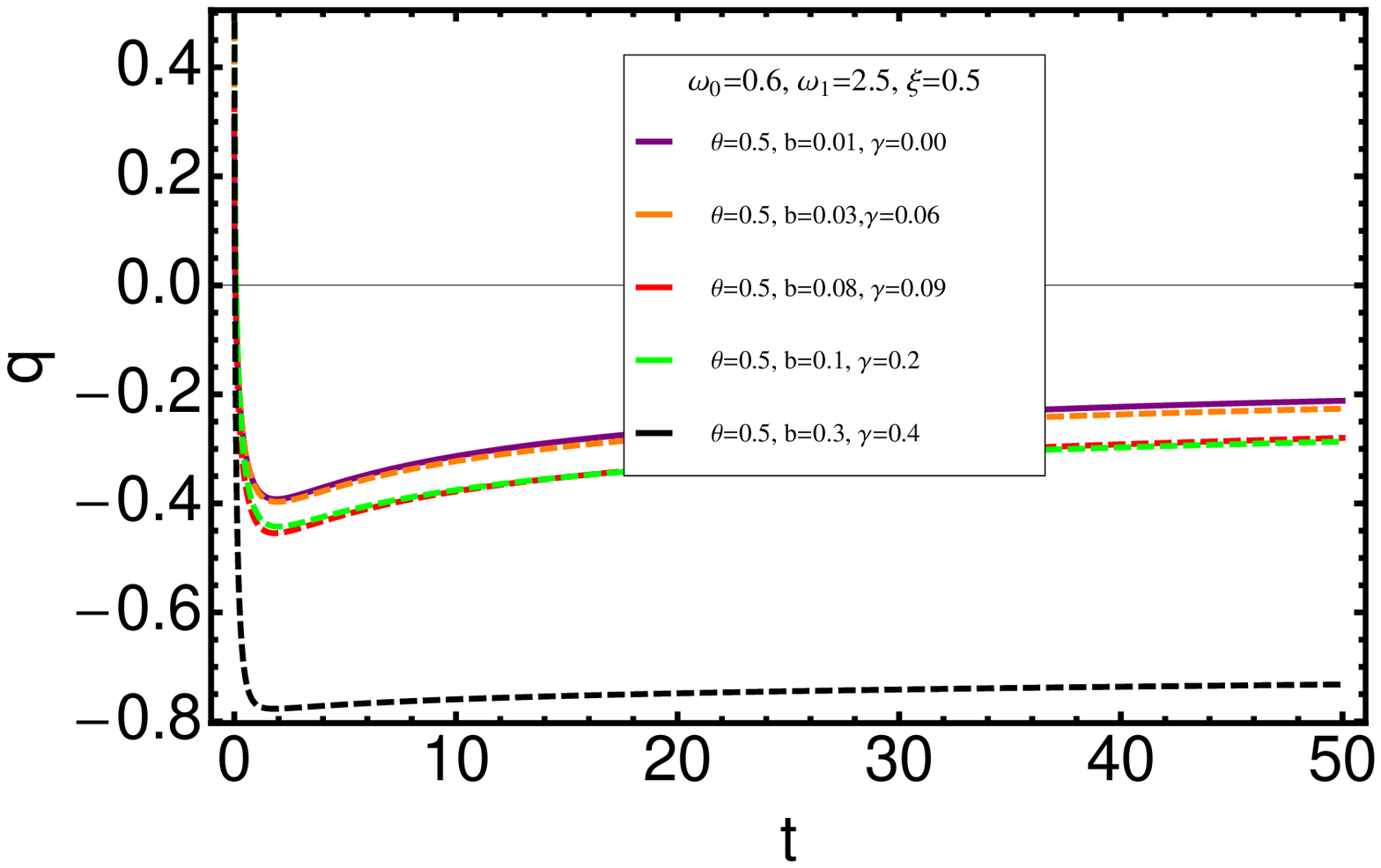} &
\includegraphics[width=50 mm]{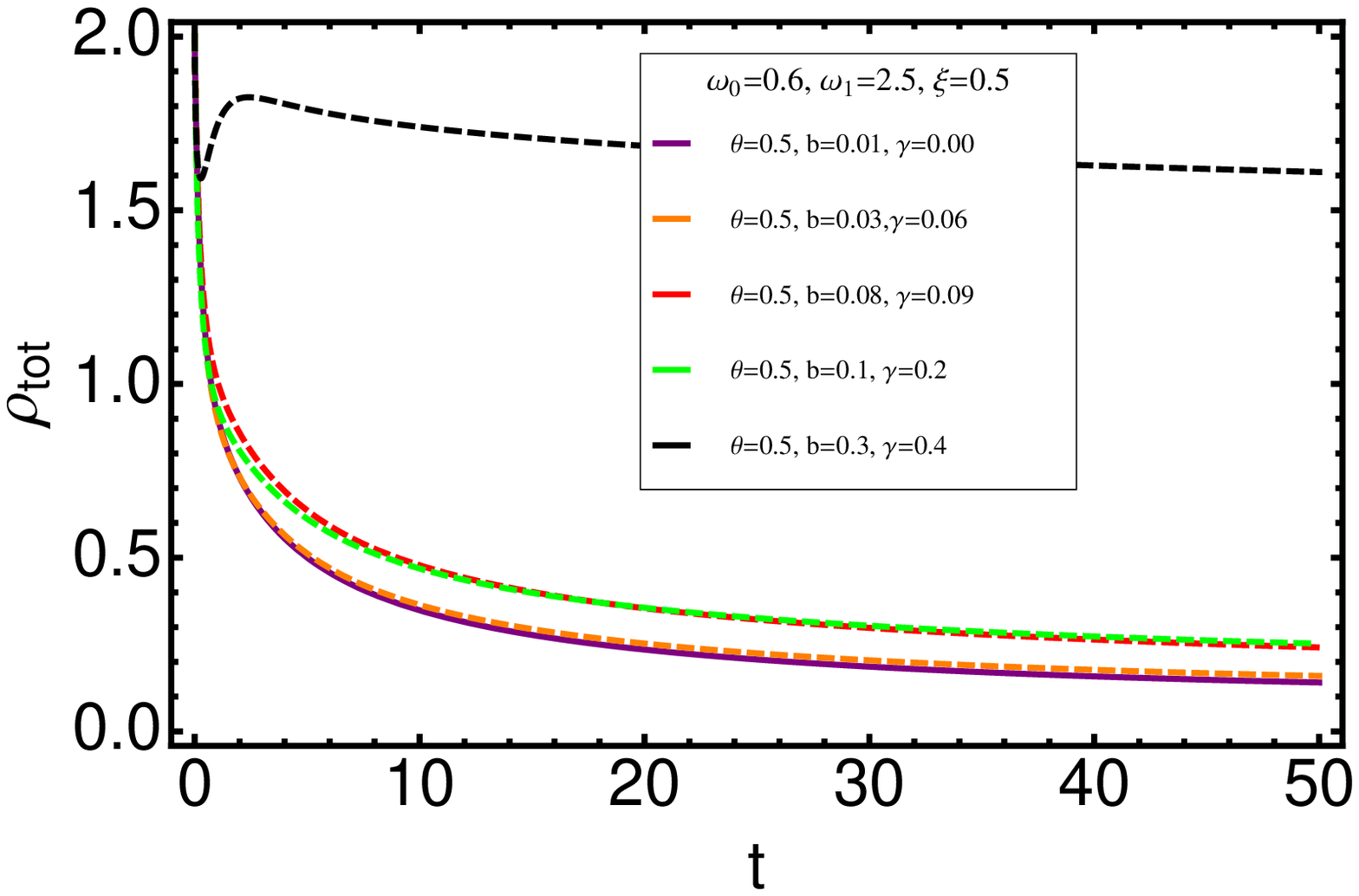}
 \end{array}$
 \end{center}
\caption{Behavior of $H$, $\omega_{tot}$, $q$ and $\rho_{tot}$ against $t$ for a Universe with interacting barotropic fluid and generalized ghost dark energy corresponding to parametrization $I$.}
 \label{fig:5}
\end{figure}

\subsubsection{Parametrization $II$}
If we use relation (13) in the equation (17) then we can study behavior of cosmological parameters $H$, $\omega_{tot}$, $q$ and $\rho_{tot}$ which illustrated in the plots of The Fig. 6.
\begin{figure}[h!]
 \begin{center}$
 \begin{array}{cccc}
\includegraphics[width=50 mm]{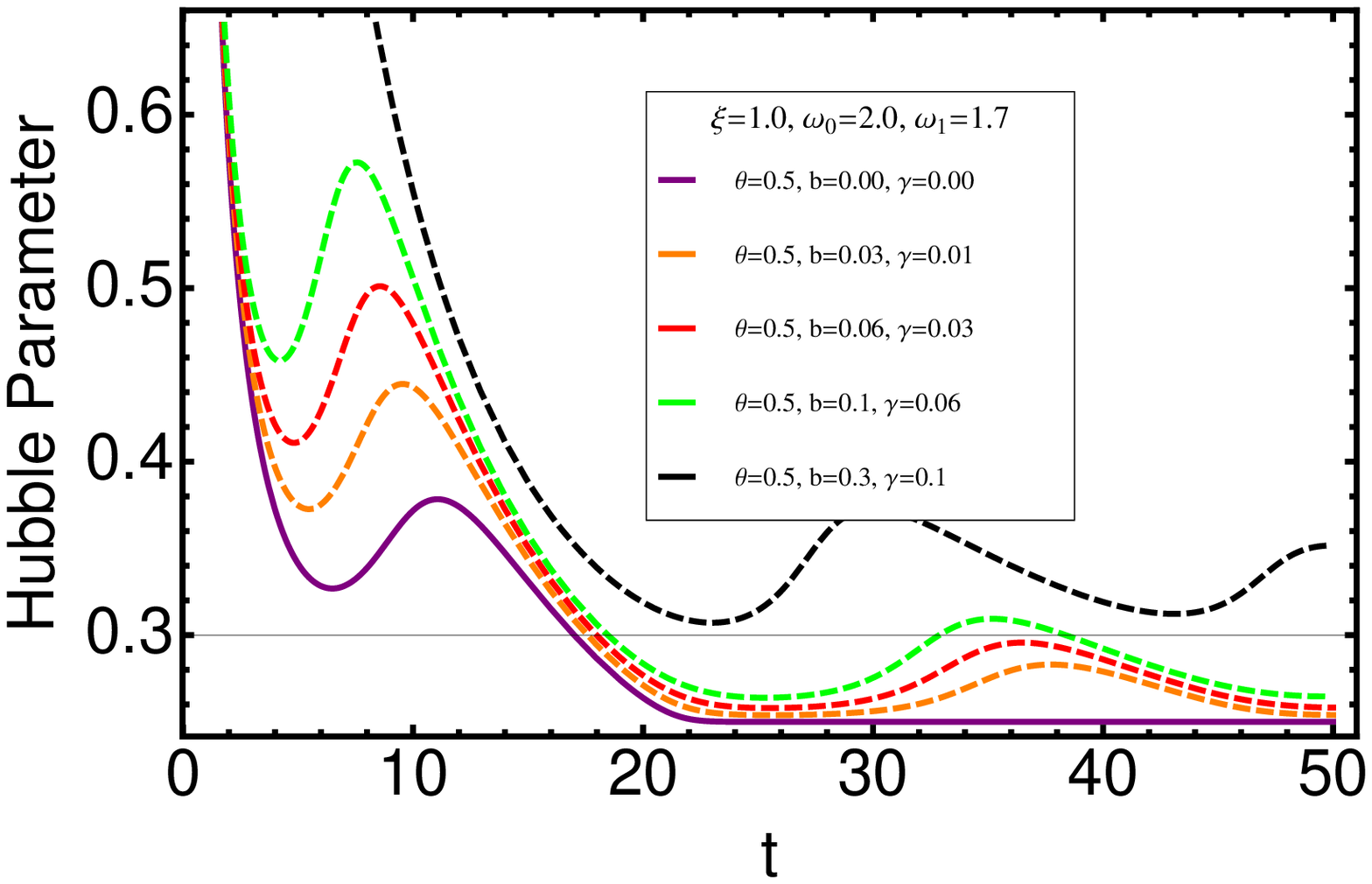} &
\includegraphics[width=50 mm]{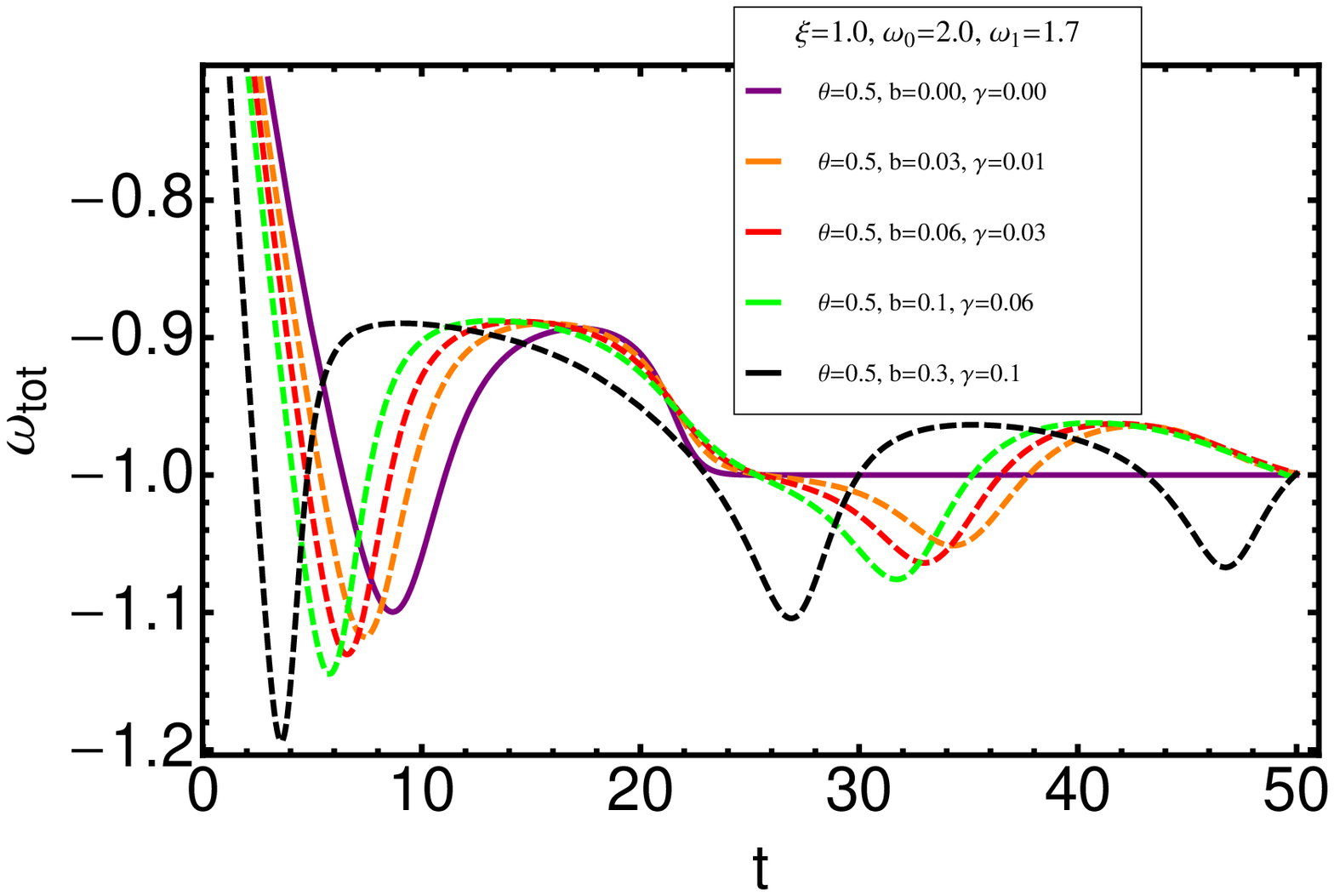}\\
\includegraphics[width=50 mm]{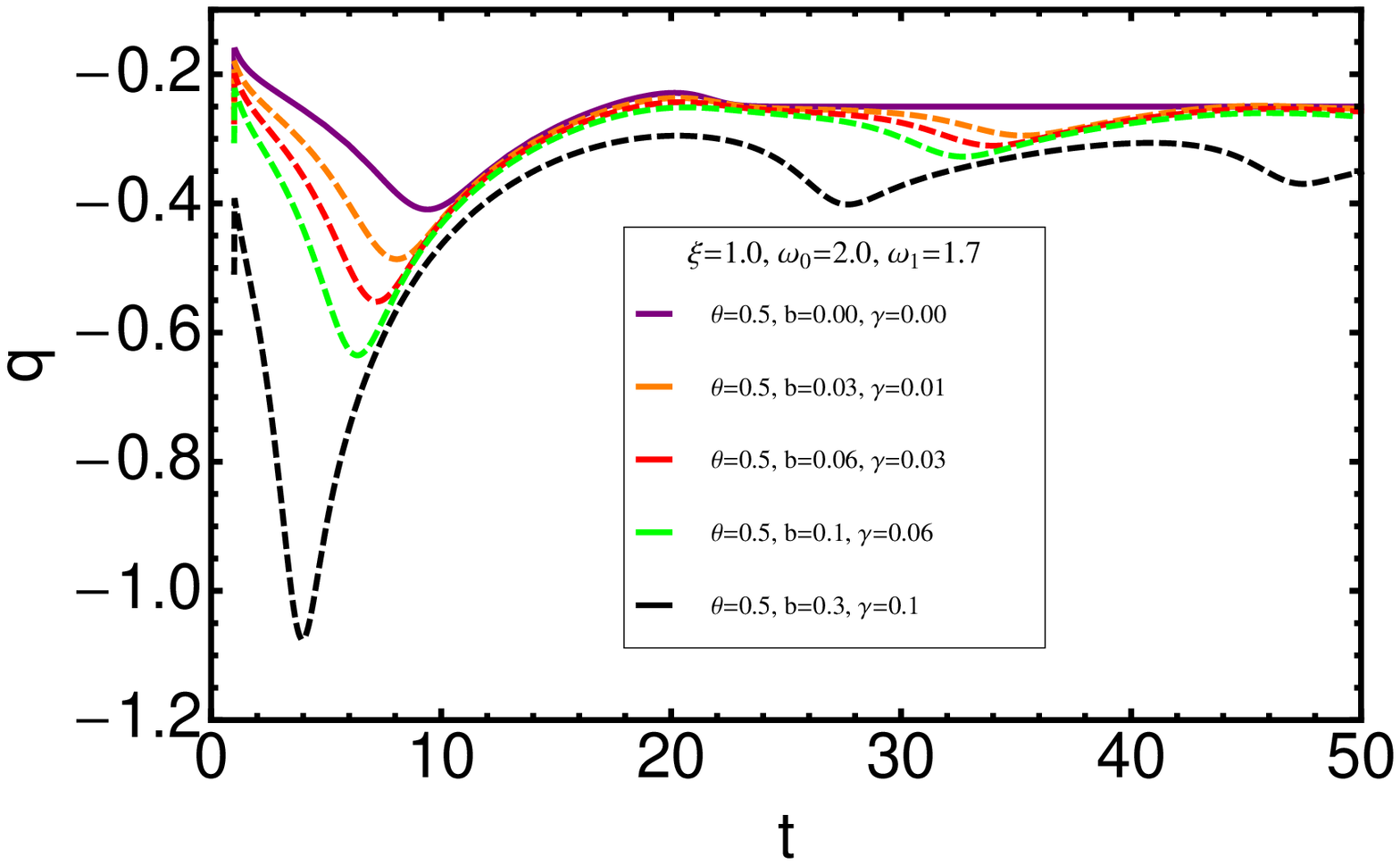} &
\includegraphics[width=50 mm]{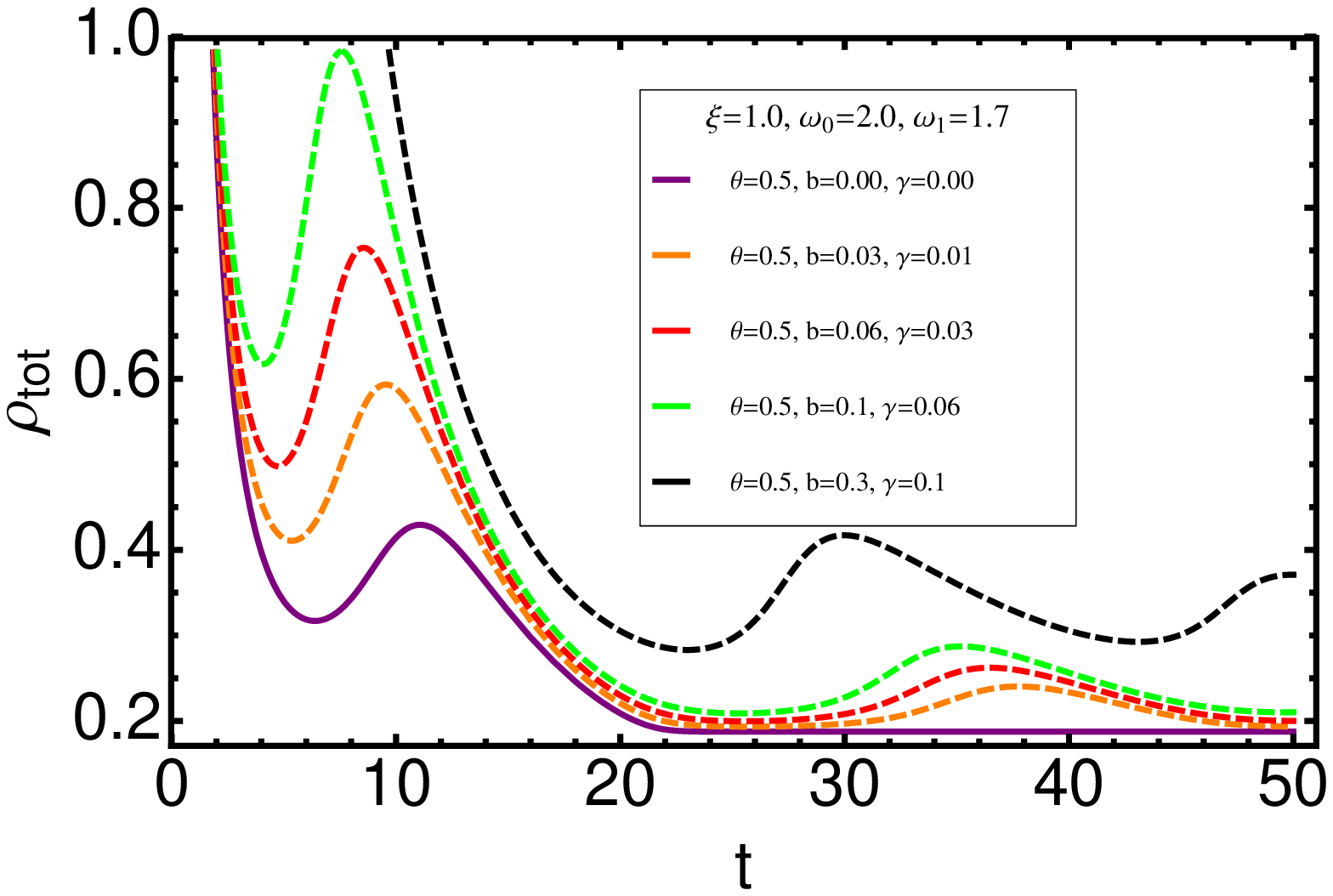}
 \end{array}$
 \end{center}
\caption{Behavior of $H$, $\omega_{tot}$, $q$ and $\rho_{tot}$ against $t$ for a Universe with interacting barotropic fluid and generalized ghost dark energy corresponding to parametrization $II$.}
 \label{fig:6}
\end{figure}

\subsection{The second model}
In the second case we consider interacting Van der Waals gas and generalized ghost dark energy model and set $\rho_{1}=\rho_{W}$ in the equations (15) and (16).
Dynamics of energy density of Van der Waals gas with a varying EoS parameter $\omega(t)$ will be described by the following equation,
\begin{equation}\label{eq:dynamicsW}
(1-\gamma)\dot{\rho}_{W}+3H\rho_{W}(1-b)+3HP_{W}-3H^{2}b(\theta+\xi H)-\gamma (\theta \dot{H}+2\xi H \dot{H})=0,
\end{equation}
Pressure of generalized ghost dark energy obtained by using the conservation equation as the following,
\begin{equation}\label{eq:pressureGDE}
P_{GGDE}=\frac{-Q-\dot{\rho}_{GD}}{3H}-\rho_{GD}.
\end{equation}
Results are different for each parametrization which illustrated in the following.
\subsubsection{Parametrization $I$}
If we use relation (11) in the equation (19) then we can study behavior of cosmological parameters $H$, $\omega_{tot}$, $q$ and $\rho_{tot}$ which illustrated in the plots of The Fig. 7.
\begin{figure}[h!]
 \begin{center}$
 \begin{array}{cccc}
\includegraphics[width=50 mm]{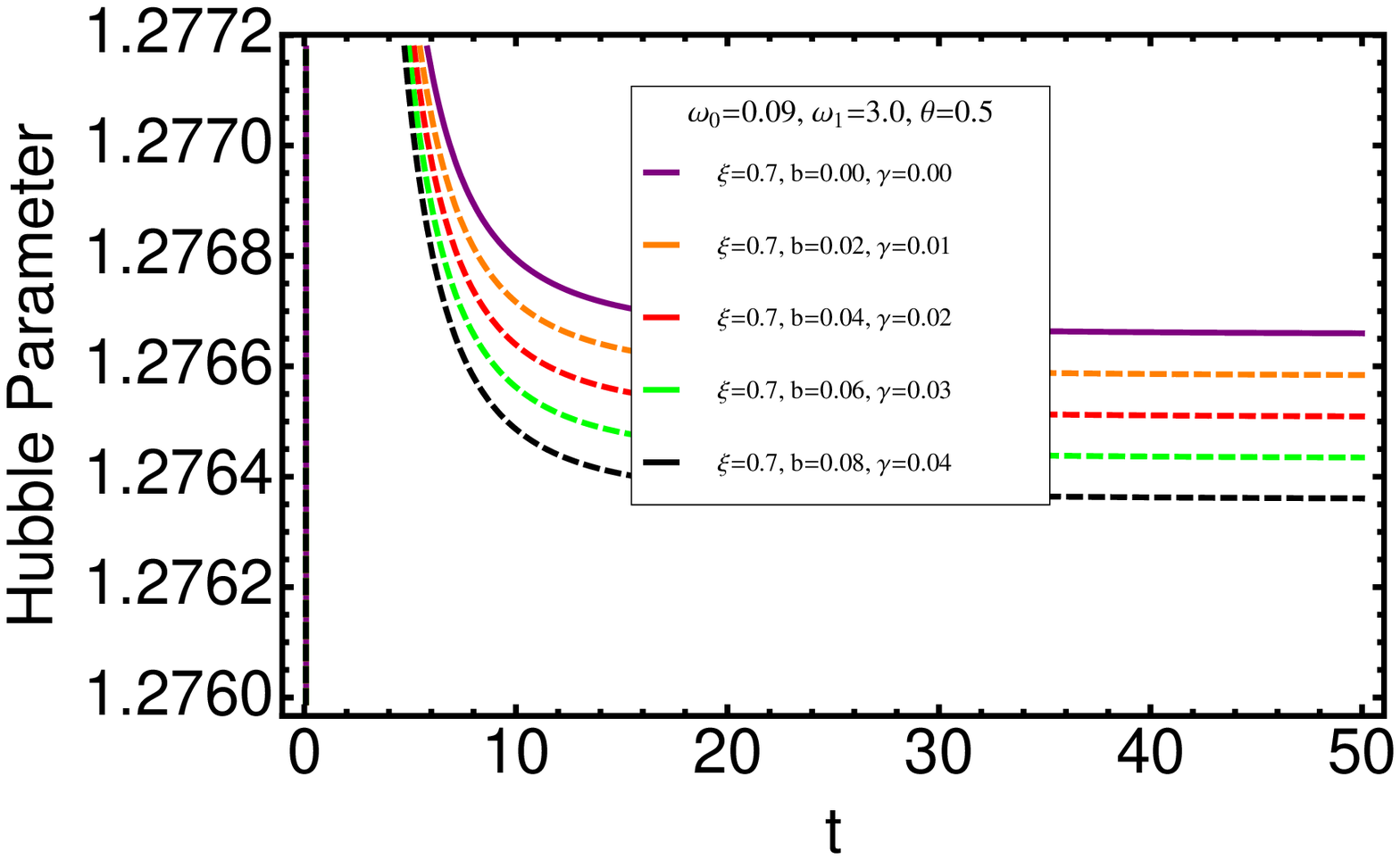} &
\includegraphics[width=50 mm]{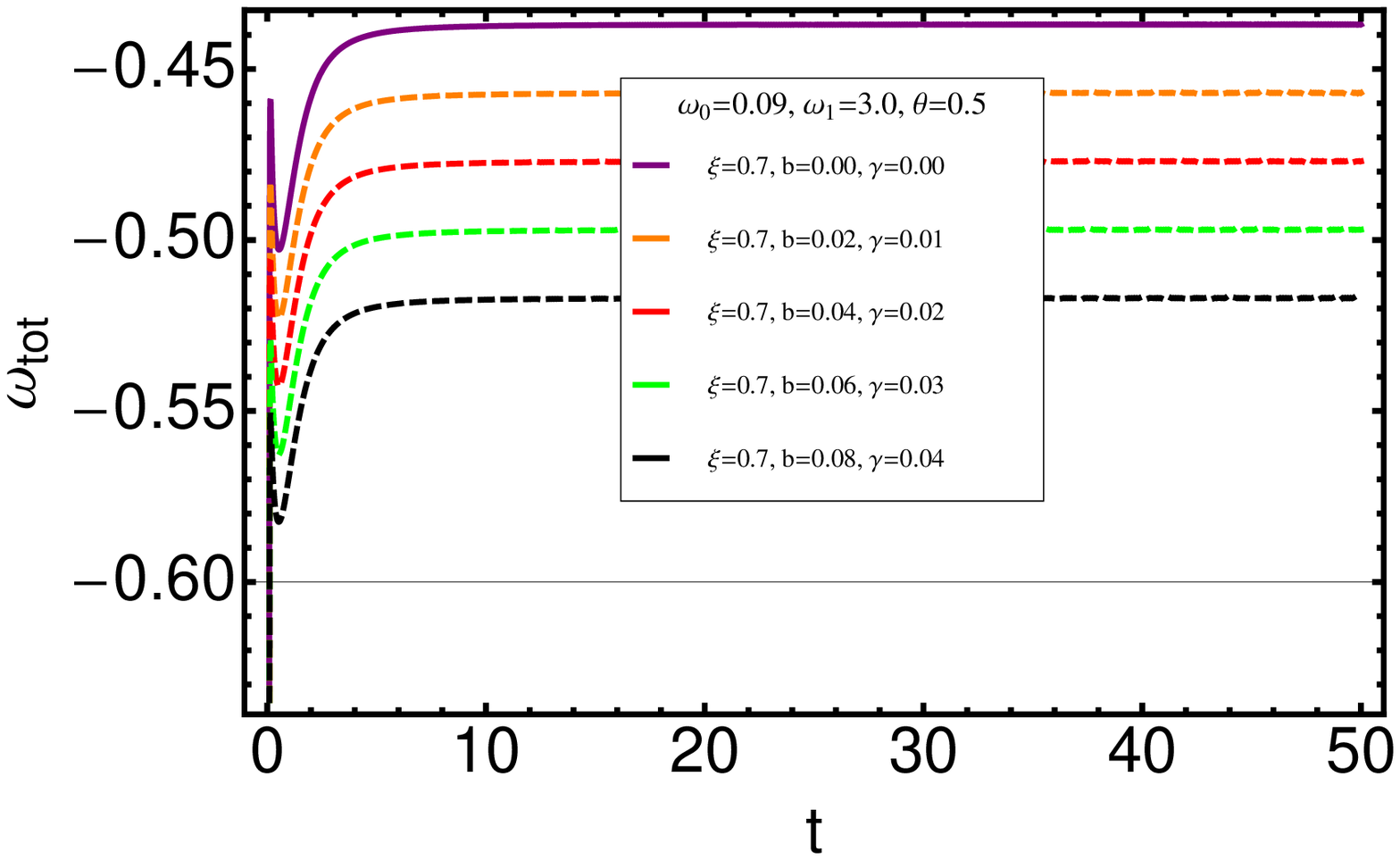}\\
\includegraphics[width=50 mm]{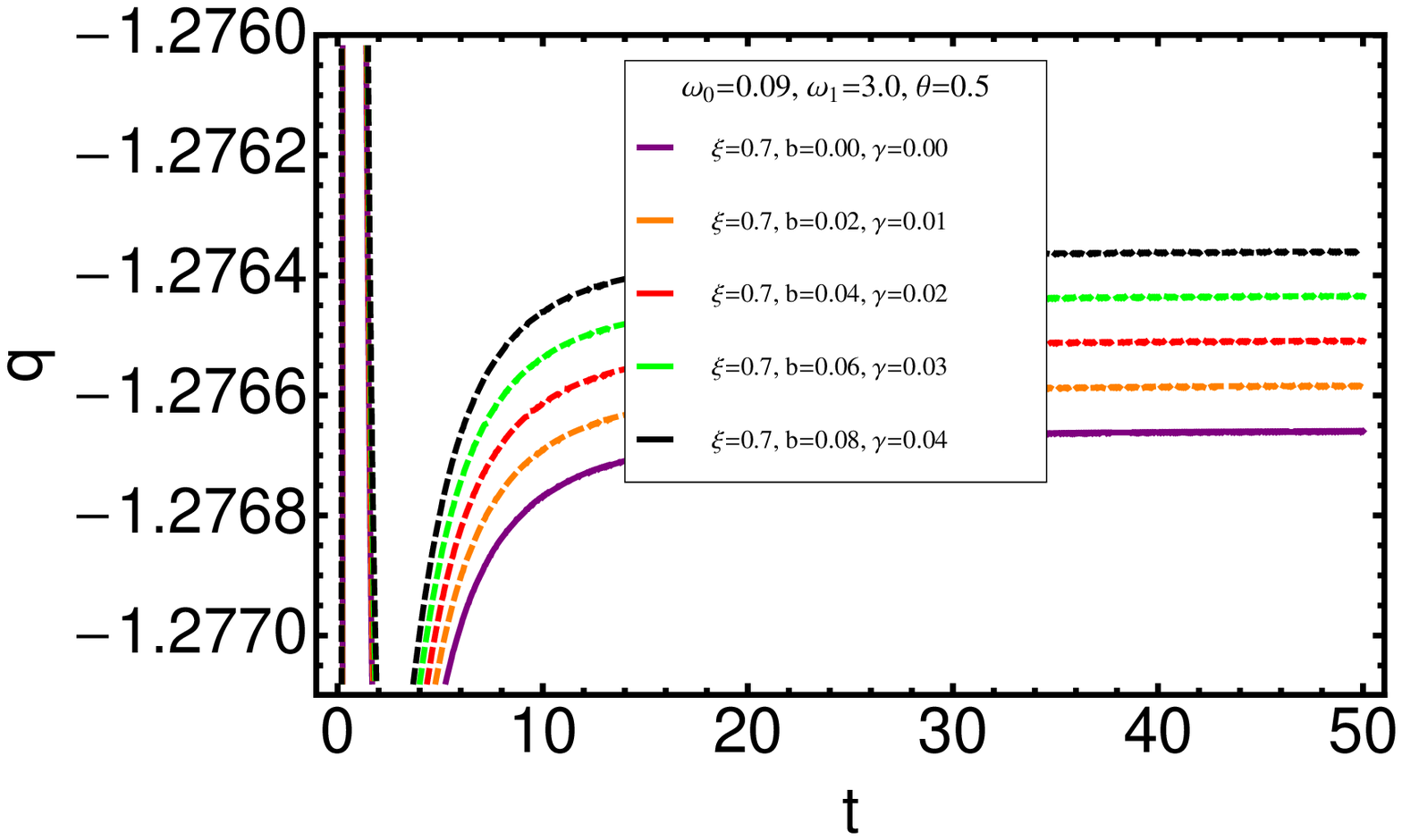} &
\includegraphics[width=50 mm]{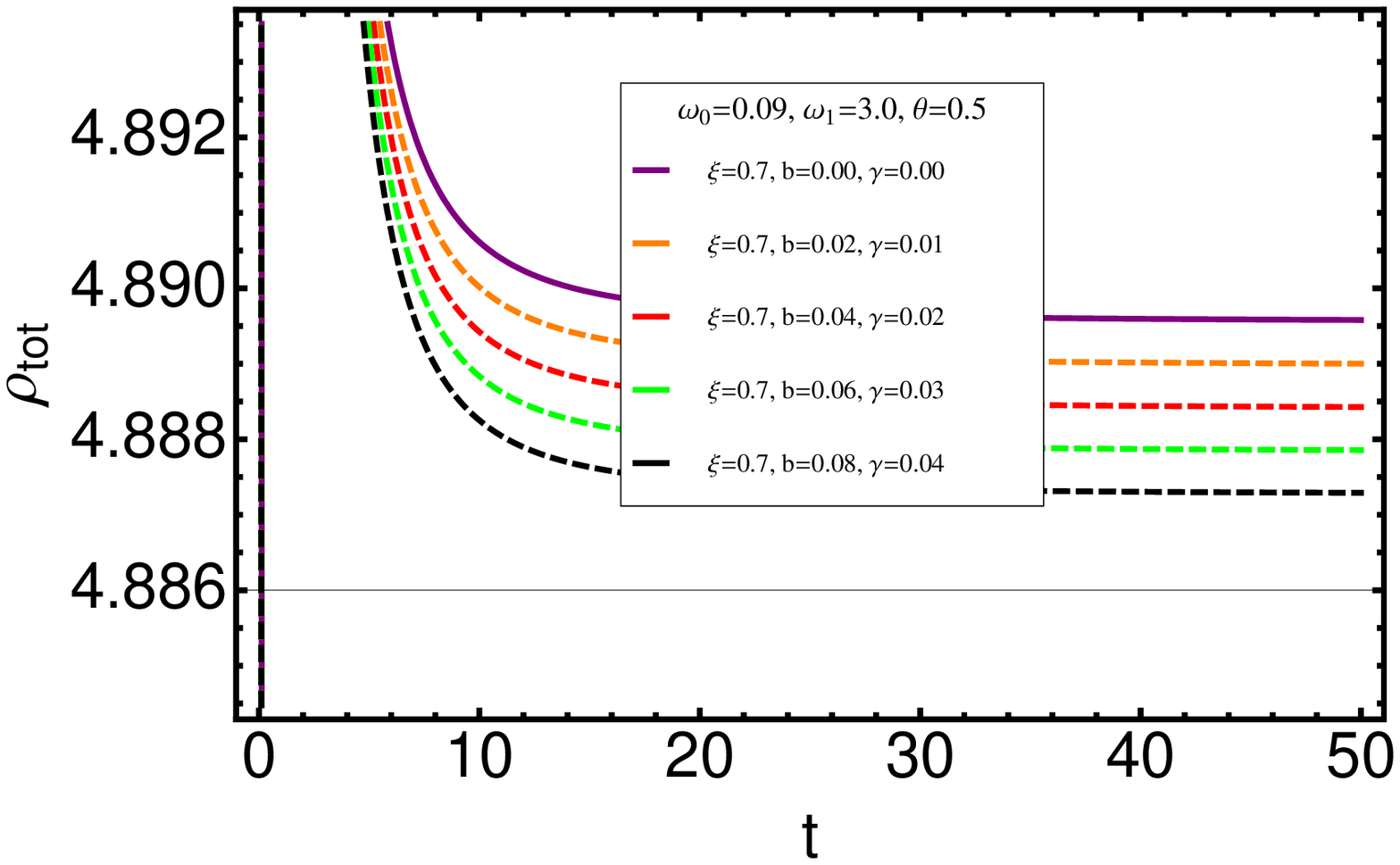}
 \end{array}$
 \end{center}
\caption{Behavior of $H$, $\omega_{tot}$, $q$ and $\rho_{tot}$ against $t$ for a Universe with interacting Van der Waals gas and generalized ghost dark energy corresponding to parametrization $I$.}
 \label{fig:7}
\end{figure}
\subsubsection{Parametrization $II$}
If we use relation (13) in the equation (19) then we can study behavior of cosmological parameters $H$, $\omega_{tot}$, $q$ and $\rho_{tot}$ which illustrated in the plots of The Fig. 8.
\begin{figure}[h!]
 \begin{center}$
 \begin{array}{cccc}
\includegraphics[width=50 mm]{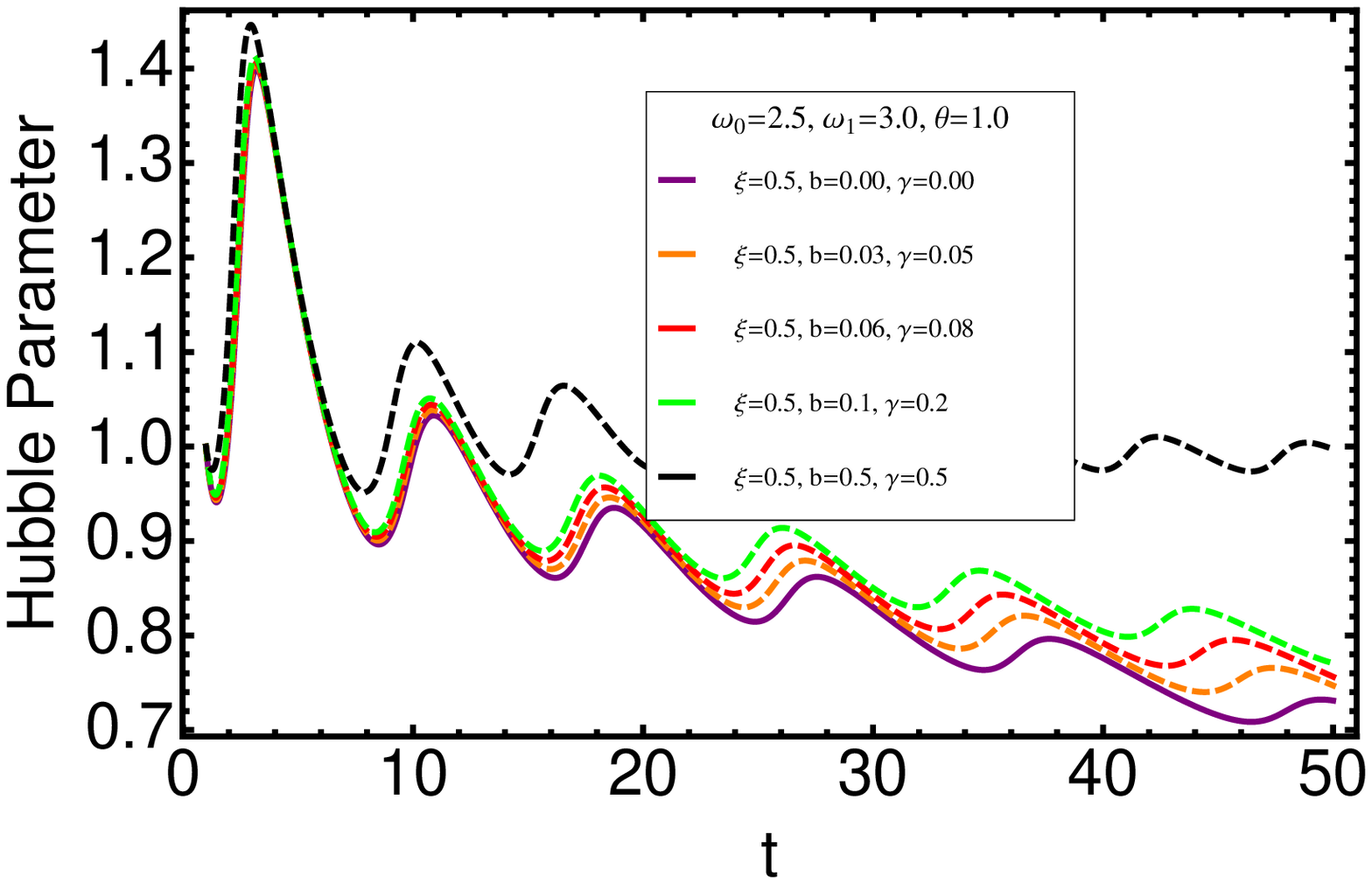} &
\includegraphics[width=50 mm]{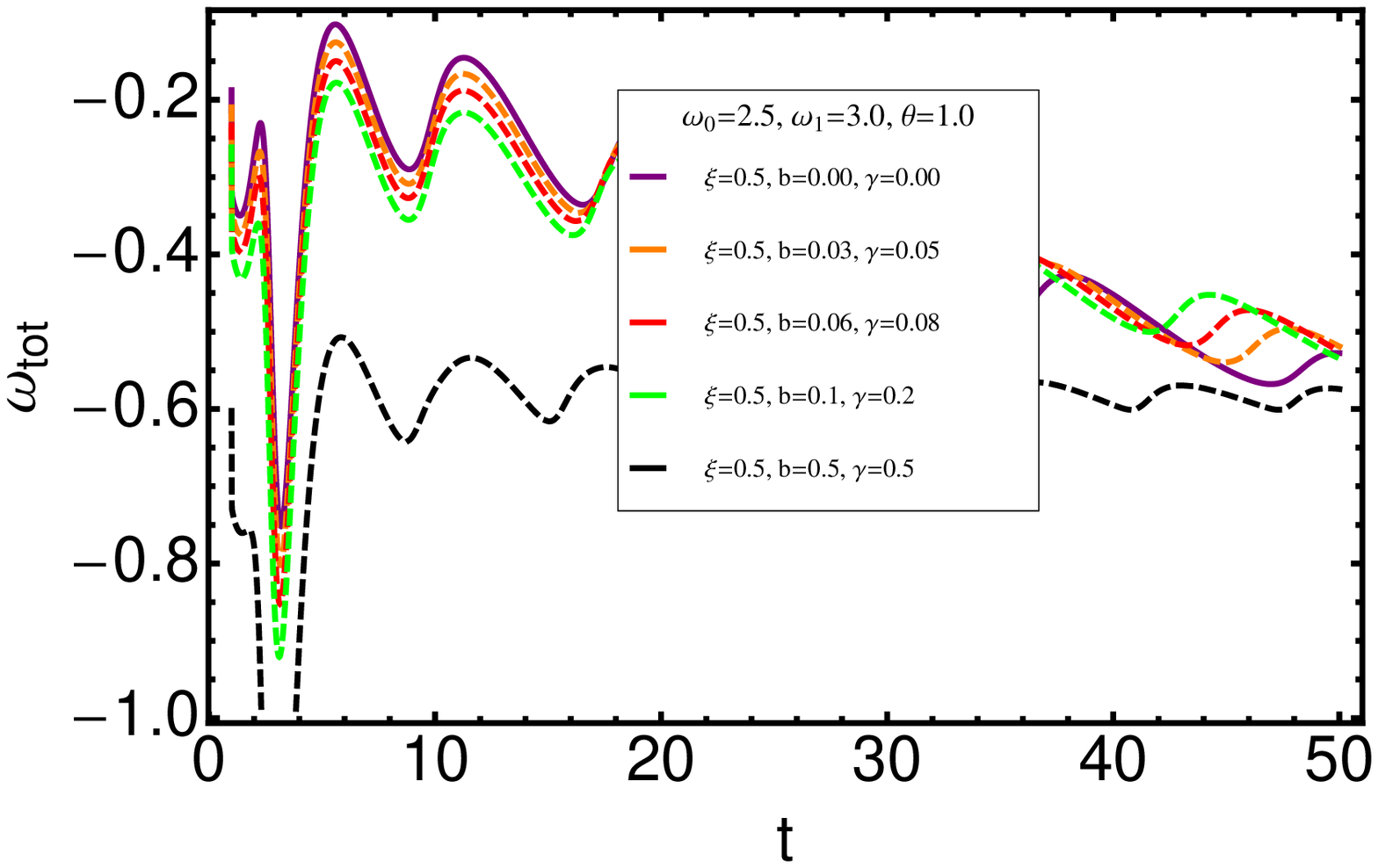}\\
\includegraphics[width=50 mm]{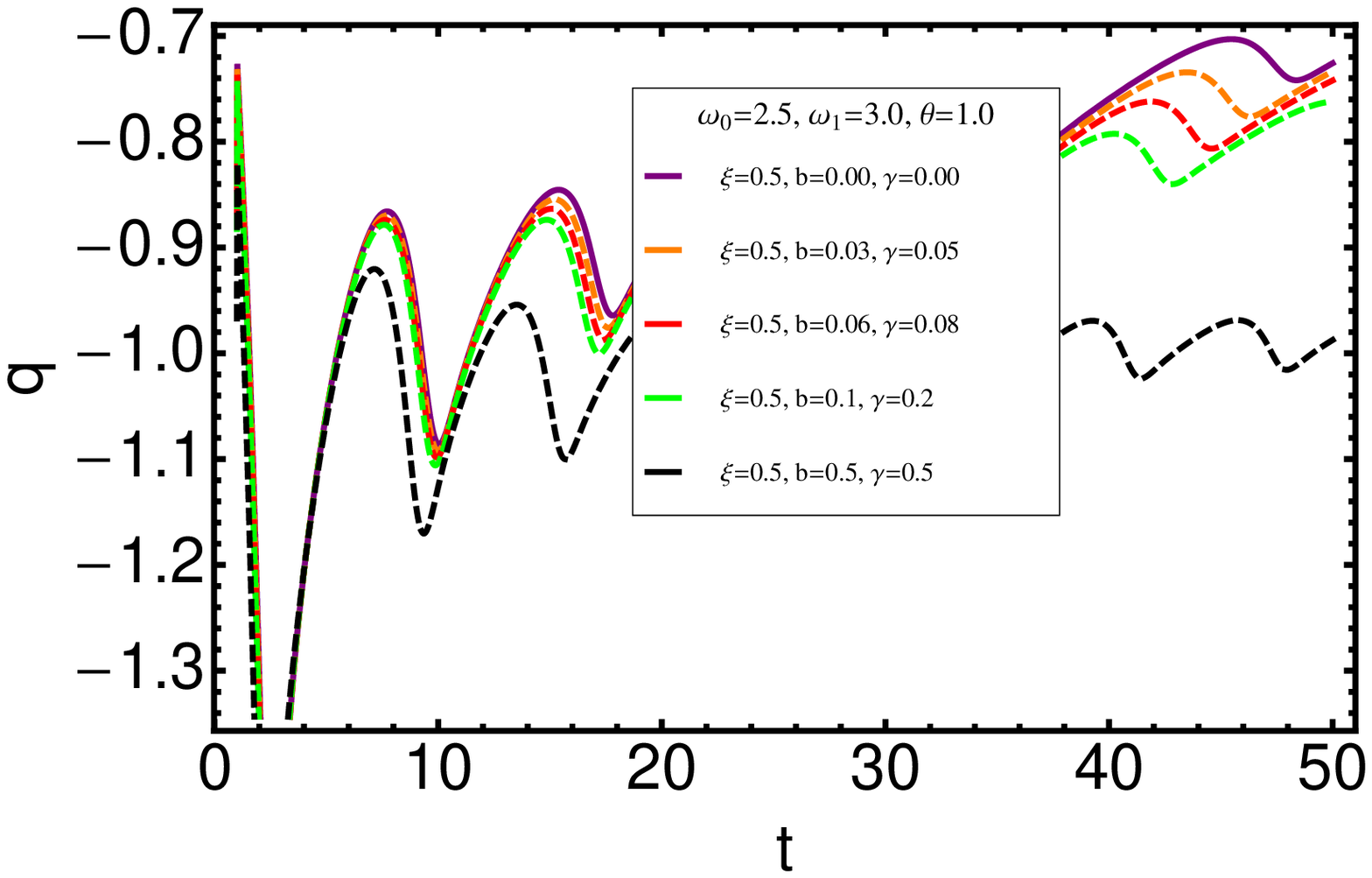} &
\includegraphics[width=50 mm]{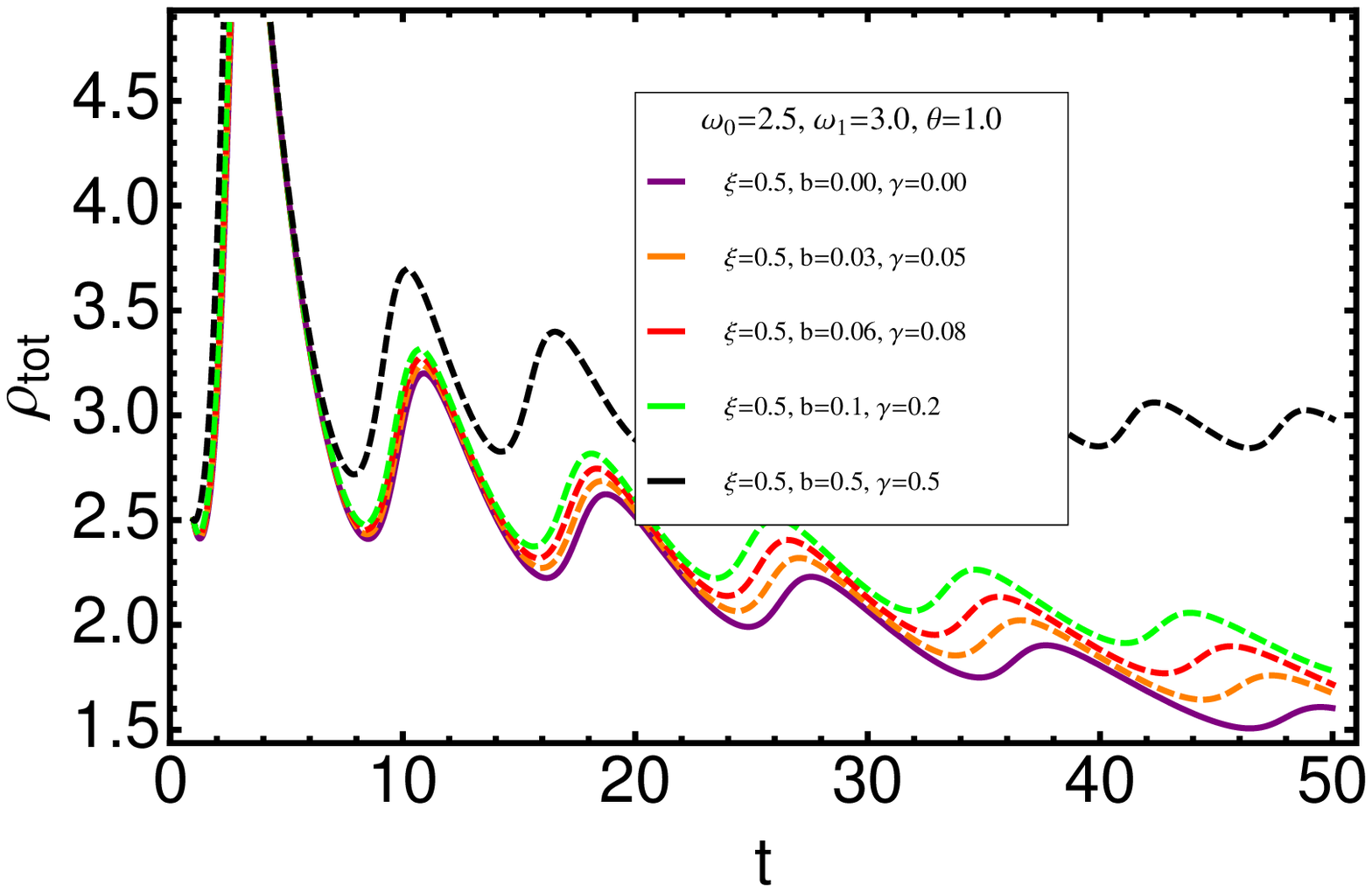}
 \end{array}$
 \end{center}
\caption{Behavior of $H$, $\omega_{tot}$, $q$ and $\rho_{tot}$ against $t$ for a Universe with interacting Van der Waals gas and generalized ghost dark energy corresponding to parametrization $II$.}
 \label{fig:8}
\end{figure}
\section{Conclusion and discussion}
In this paper we assumed new models for the Universe based on interacting two-fluid component. In the first model we assumed that barotropic fluid, as the first component, interact with generalized ghost dark energy, as the second component. In the second model we assumed that Van der Waals gas, as the first component, interact with generalized ghost dark energy, as the second component. In both models we used two different time-dependent parametrization. Therefore, we have four different situations: The first model with the first parametrization (M$I$ P$I$), The second model with the first parametrization (M$I$ P$II$), The second model with the first parametrization (M$II$ P$I$), The second model with the second parametrization (M$II$ P$II$). First of all we studied cosmological parameters in the case of non-interacting components (Figs. 1-4), and then we studied that in the case of interacting components (Figs. 5-8).\\
Fig. 1 studied graphically the case of non-interacting M$I$ P$I$ and drawn $H$, $\omega_{tot}$ and $q$. We found $H$ decreasing function of time which its value increased by $\omega_{0}$ and $\omega_{1}$. The value of $\omega_{tot}$ yields a constant at the late time. Closest value to -1 obtained for $\omega_{0}=1.5$ and $\omega_{1}=0.7$. Deceleration parameter of this case yields to zero at the late time.\\
Fig. 2 studied graphically the case of non-interacting M$I$ P$II$ and drawn $H$, $\omega_{tot}$ and $q$. We found $H$ increasing function of time which its value decreased by $\omega_{0}$ and $\omega_{1}$. The value of $\omega_{tot}$ periodic picture which oscillate around the origin. Deceleration parameter of this case is decreasing function at the late time and is increasing function at the initial time.\\
Fig. 3 studied graphically the case of non-interacting M$II$ P$I$ and drawn $H$, $\omega_{tot}$ and $q$. We found $H$ decreasing function of time which its value increased by $\omega_{0}$ and $\omega_{1}$. The value of $\omega_{tot}$ yields a constant near zero at the late time. Deceleration parameter of this case yields to -1 at the late time.\\
Fig. 4 studied graphically the case of non-interacting M$II$ P$II$ and drawn $H$, $\omega_{tot}$ and $q$. We found $H$ quasi-periodic decreasing function of time which its value decreased by $\omega_{0}$ and $\omega_{1}$ at the late time. Similar behavior can be seen for the $\omega_{tot}$. Deceleration parameter of this case has also quasi-periodic picture and yields to a constant at the late time.\\
Fig. 5 studied graphically the case of interacting M$I$ P$I$ and drawn $H$, $\omega_{tot}$, $q$ and $\rho_{tot}$. We found $H$ decreasing function of time which its value increased by interaction parameter. The value of $\omega_{tot}$ yields to -1 at the late time. Closest value to -1 obtained for higher value of interaction parameters. Deceleration parameter of this case yields a constant at the late time. Finally, the total density of this case decreased by time and increased by interaction parameters. For the special values of interaction parameters may be have increasing density at initial time.\\
Fig. 6 studied graphically the case of interacting M$I$ P$II$ and drawn $H$, $\omega_{tot}$, $q$ and $\rho_{tot}$. We found $H$ decreasing function of time which its value increased by interaction parameters. The value of $\omega_{tot}$ has quasi-periodic picture which oscillate around the -1 and yields to that at the late time. Deceleration parameter of this case is also yields to a constant at the late time, its value decreased by increasing interaction parameters.\\
Fig. 7 studied graphically the case of interacting M$II$ P$I$ and drawn $H$, $\omega_{tot}$, $q$ and $\rho_{tot}$. We found $H$ decreasing function of time which its value decreased by interaction parameters. The value of $\omega_{tot}$ yields a constant at the late time. Deceleration parameter of this case yields to -1.2 at the late time. The total density of this case decreased by time and also decreased by interaction parameters.\\
Fig. 8 studied graphically the case of interacting M$II$ P$II$ and drawn $H$, $\omega_{tot}$, $q$ and $\rho_{tot}$. We found $H$ quasi-periodic decreasing function of time which its value increased by interaction parameters. $\omega_{tot}$ has also quasi-periodic picture which yields to a constant at the late time. Deceleration parameter of this case has also quasi-periodic decreasing behavior and decreased by interaction parameters. Finally behavior of total density is similar to $\omega_{tot}$.\\
Now, we should compare our result with observational data. According to the $1\sigma$ level from $H(z)$ data $q\approx-0.3$ and $H_{0}=68.43\pm2.8 \frac{Km}{s Mpc}$ [29]. On the other hand from data of SNe Ia we have $q\approx-0.43$, $-1.67<\omega<-0.62$ and $H_{0}=69.18\pm0.55 \frac{Km}{s Mpc}$ [29, 30]. Also joint test using $H(z)$ and SNe Ia give $-0.39\leq q\leq-0.29$ and $H_{0}=68.93\pm0.53 \frac{Km}{s Mpc}$ [29]. Recent astronomical data based on anew infrared camera on the HST gives $H_{0}=73.8\pm2.4 \frac{Km}{s Mpc}$ [31]. The other prob using galactic clusters data suggest $H_{0}=67\pm3.2 \frac{Km}{s Mpc}$ [32]. Finally, $\Lambda$CDM model suggests $q\rightarrow-1$ and the best fitted parameters of the Ref. [33] say that $q=-0.64$.\\
We would like to use above observational data to find appropriate models used in this paper. According to the Fig.1 the M$I$ P$I$ is far from any observation. According to the Fig. 2 the M$I$ P$II$ with $\omega_{0}=0.7$ and $\omega_{1}=2$ more agree with joint test using $H(z)$ and SNe Ia. According to the Fig. 3 the M$II$ P$I$ is far from observational data just $\omega_{0}=0.01$ and $\omega_{1}=0.75$ recover relatively $\Lambda$CDM model. According to the Fig. 4 the M$II$ P$II$ may be close to the best fitted parameters of the Ref. [33] for $\omega_{0}=2.5$ and $\omega_{1}=3$ or $\Lambda$CDM model for $\omega_{0}=0.5$ and $\omega_{1}=1.5$. Therefore, the M$II$ P$II$ (combination of Van der Waals gas and generalized ghost dark energy with $\omega(t)=\omega_{0}\cos(tH)+\omega_{1}t\frac{\dot{H}}{H}$) may be the best model in the case of non-interaction.\\
We expect that a model including interaction will be the best model of this paper. Fig. 5 suggest that M$I$ P$I$ recovers the $1\sigma$ level of $H(z)$ data for $\omega_{0}=0.6$, $\omega_{1}=2.5$, $\xi=\theta=0.5$, $b=0.1$ and $\gamma=0.2$. (green line of the Fig. 5). Black line of the Fig. 6 more coincides with the SNe Ia data therefore the M$I$ P$II$ is also good model. Plots of the Fig. 7 suggest that M$II$ P$I$ has no agreement with any observational data. Finally, plots Fig 8 show that the best fitted parameters of the Ref. [33] may be recovered by choosing $\omega_{0}=2.5$, $\omega_{1}=3$, $\xi=0.5$, $\theta=1$, $b=0$ and $\gamma=0$ which is again non-interacting case.\\
Therefore we conclude that the M$I$ P$II$ may be the best model to describe Universe within the aim of this paper. It means that combination of barotropic fluid with linear time dependent equation of state $\omega(t)=\omega_{0}\cos(tH)+\omega_{1}t\frac{\dot{H}}{H}$ and generalized ghost dark energy may be appropriate model of our Universe. However the model including Van der Waals gas is relatively good model. We also conclude that parametrization $I$ dose not work good.

\end{document}